\documentclass[twocolumn]{aastex701}
\usepackage{amsmath}

\newcommand{\Zism}{Z_{\mathrm{ISM}}}
\newcommand{\ms}{M_*}
\newcommand{\dms}{\dot{M}_*}
\newcommand{\mh}{M_h}
\newcommand{\msret}{M_*^{\mathrm{ret}}}
\newcommand{\msobs}{M_*^{\mathrm{obs}}}
\newcommand{\msfmd}{M_*^{\mathrm{fmd}}}
\newcommand{\sfr}{\mathrm{SFR}}
\newcommand{\sfrobs}{\mathrm{SFR}^{\mathrm{obs}}}
\newcommand{\gyr}{\mathrm{Gyr}}
\newcommand{\Rscale}{R_{\mathrm{scale}}}
\newcommand{\rscale}{r_{\mathrm{scale}}}
\newcommand{\msun}{M_\odot}
\newcommand{\sat}{\texttt{SAT}}
\newcommand{\fone}{\texttt{F}\emph{obs}}
\newcommand{\ftwo}{\texttt{F}\emph{fmd}}
\newcommand{\fthree}{\texttt{F}\emph{sfh}}
\newcommand{\tausfh}{\tau_{\mathrm{sfh}}}
\newcommand{\ssfr}{s\mathrm{SFR}}
\newcommand{\tauin}{\tau_{\mathrm{infall}}}
\newcommand{\fin}{f_{\mathrm{inflow}}}
\newcommand{\powexp}{\texttt{powexp}}
\newcommand{\modot}{\dot{M}_{\mathrm{O}}}
\newcommand{\zo}{{Z}_{\mathrm{O}}}
\newcommand{\zoeq}{{Z}_{\mathrm{O}, \mathrm{eq}}}
\newcommand{\zoinf}{{Z}_{\mathrm{inf}}}

\newcommand{\mocc}{{m}^{\mathrm{cc}}_{\mathrm{O}}}
\newcommand{\mo}{{M}_{\mathrm{O}}}
\newcommand{\mgas}{{M}_{\mathrm{gas}}}
\newcommand{\dmginf}{\dot{M}_{\mathrm{inf}}}
\newcommand{\dmgout}{\dot{M}_{\mathrm{out}}}
\newcommand{\frec}{f_{\mathrm{rec}}}
\newcommand{\finf}{f_{\mathrm{inflow}}}
\newcommand{\ficm}{f_{\mathrm{inflow}}^{\mathrm{ICM}}}
\newcommand{\figm}{f_{\mathrm{inflow}}^{\mathrm{IGM}}}
\newcommand{\dd}{\mathrm{d}}
\newcommand{\mm}{\mathrm{m}}

\newcommand{\hscale}{h_{\mathrm{scale}}}

\begin{document}

\title{Satellite Metallicity Enhancement I: Suppressed Star Formation, Stellar Mass Loss, and Enriched Inflow of DESI and EAGLE Galaxies around Massive Clusters}

\author[]{Yuanye Lin}
\affiliation{State Key Laboratory of Dark Matter Physics \& Tsung-Dao Lee
Institute, Shanghai Jiao Tong University, Shanghai, 200240, China.}
\affiliation{Department of Astronomy, School of Physics and Astronomy, Shanghai Jiao Tong University, Shanghai 200240, China}
\email{ythylyy@sjtu.edu.cn}

\author[]{Ying Zu}
\affiliation{State Key Laboratory of Dark Matter Physics \& Tsung-Dao Lee
Institute, Shanghai Jiao Tong University, Shanghai, 200240, China.}
\affiliation{Department of Astronomy, School of Physics and Astronomy, Shanghai Jiao Tong University, Shanghai 200240, China}
\email[show]{yingzu@sjtu.edu.cn}

\author[]{J.~Aguilar}
\affiliation{Lawrence Berkeley National Laboratory, 1 Cyclotron Road, Berkeley, CA 94720, USA}
\email{jaguilar@lbl.gov}

\author[]{S.~Ahlen}
\affiliation{Department of Physics, Boston University, 590 Commonwealth Avenue, Boston, MA 02215 USA}
\email{ahlen@bu.edu}

\author[]{D.~Brooks}
\affiliation{Department of Physics \& Astronomy, University College London, Gower Street, London, WC1E 6BT, UK}
\email{david.brooks@ucl.ac.uk}

\author[]{T.~Claybaugh}
\affiliation{Lawrence Berkeley National Laboratory, 1 Cyclotron Road, Berkeley, CA 94720, USA}
\email{tmclaybaugh@lbl.gov}

\author[]{A.~Cuceu}
\affiliation{Lawrence Berkeley National Laboratory, 1 Cyclotron Road, Berkeley, CA 94720, USA}
\email{acuceu@lbl.gov}

\author[]{A.~de la Macorra}
\affiliation{Instituto de F\'{\i}sica, Universidad Nacional Aut\'{o}noma de M\'{e}xico,  Circuito de la Investigaci\'{o}n Cient\'{\i}fica, Ciudad Universitaria, Cd. de M\'{e}xico  C.~P.~04510,  M\'{e}xico}
\email{macorra@fisica.unam.mx}

\author[]{A.~Font-Ribera}
\affiliation{Instituci\'{o} Catalana de Recerca i Estudis Avan\c{c}ats, Passeig de Llu\'{\i}s Companys, 23, 08010 Barcelona, Spain}
\affiliation{Institut de F\'{i}sica d’Altes Energies (IFAE), The Barcelona Institute of Science and Technology, Edifici Cn, Campus UAB, 08193, Bellaterra (Barcelona), Spain}
\email{afont@ifae.es}

\author[]{J.~E.~Forero-Romero}
\affiliation{Departamento de F\'isica, Universidad de los Andes, Cra. 1 No. 18A-10, Edificio Ip, CP 111711, Bogot\'a, Colombia}
\affiliation{Observatorio Astron\'omico, Universidad de los Andes, Cra. 1 No. 18A-10, Edificio H, CP 111711 Bogot\'a, Colombia}
\email{je.forero@uniandes.edu.co}

\author[]{Satya~{Gontcho A Gontcho}}
\affiliation{University of Virginia, Department of Astronomy, Charlottesville, VA 22904, USA}
\email{satya@virginia.edu}

\author[]{G.~Gutierrez}
\affiliation{Fermi National Accelerator Laboratory, PO Box 500, Batavia, IL 60510, USA}
\email{gaston@fnal.gov}

\author[]{R.~Joyce}
\affiliation{NSF NOIRLab, 950 N. Cherry Ave., Tucson, AZ 85719, USA}
\email{richard.joyce@noirlab.edu}

\author[]{M.~Landriau}
\affiliation{Lawrence Berkeley National Laboratory, 1 Cyclotron Road, Berkeley, CA 94720, USA}
\email{mlandriau@lbl.gov}

\author[]{L.~Le~Guillou}
\affiliation{Sorbonne Universit\'{e}, CNRS/IN2P3, Laboratoire de Physique Nucl\'{e}aire et de Hautes Energies (LPNHE), FR-75005 Paris, France}
\email{llg@lpnhe.in2p3.fr}

\author[]{A.~Meisner}
\affiliation{NSF NOIRLab, 950 N. Cherry Ave., Tucson, AZ 85719, USA}
\email{aaron.meisner@noirlab.edu}

\author[]{R.~Miquel}
\affiliation{Instituci\'{o} Catalana de Recerca i Estudis Avan\c{c}ats, Passeig de Llu\'{\i}s Companys, 23, 08010 Barcelona, Spain}
\affiliation{Institut de F\'{i}sica d’Altes Energies (IFAE), The Barcelona Institute of Science and Technology, Edifici Cn, Campus UAB, 08193, Bellaterra (Barcelona), Spain}
\email{rmiquel@ifae.es}

\author[]{J.~Moustakas}
\affiliation{Department of Physics and Astronomy, Siena University, 515 Loudon Road, Loudonville, NY 12211, USA}
\email{jmoustakas@siena.edu}

\author[]{W.~J.~Percival}
\affiliation{Department of Physics and Astronomy, University of Waterloo, 200 University Ave W, Waterloo, ON N2L 3G1, Canada}
\affiliation{Perimeter Institute for Theoretical Physics, 31 Caroline St. North, Waterloo, ON N2L 2Y5, Canada}
\affiliation{Waterloo Centre for Astrophysics, University of Waterloo, 200 University Ave W, Waterloo, ON N2L 3G1, Canada}
\email{will.percival@uwaterloo.ca}

\author[]{F.~Prada}
\affiliation{Instituto de Astrof\'{i}sica de Andaluc\'{i}a (CSIC), Glorieta de la Astronom\'{i}a, s/n, E-18008 Granada, Spain}
\email{fprada@iaa.es}

\author[]{I.~P\'erez-R\`afols}
\affiliation{Departament de F\'isica, EEBE, Universitat Polit\`ecnica de Catalunya, c/Eduard Maristany 10, 08930 Barcelona, Spain}
\email{ignasi.perez.rafols@upc.edu}

\author[]{G.~Rossi}
\affiliation{Department of Physics and Astronomy, Sejong University, 209 Neungdong-ro, Gwangjin-gu, Seoul 05006, Republic of Korea}
\email{graziano@sejong.ac.kr}

\author[]{E.~Sanchez}
\affiliation{CIEMAT, Avenida Complutense 40, E-28040 Madrid, Spain}
\email{eusebio.sanchez@ciemat.es}

\author[]{D.~Schlegel}
\affiliation{Lawrence Berkeley National Laboratory, 1 Cyclotron Road, Berkeley, CA 94720, USA}
\email{djschlegel@lbl.gov}

\author[]{J.~Silber}
\affiliation{Lawrence Berkeley National Laboratory, 1 Cyclotron Road, Berkeley, CA 94720, USA}
\email{jhsilber@lbl.gov}

\author[]{D.~Sprayberry}
\affiliation{NSF NOIRLab, 950 N. Cherry Ave., Tucson, AZ 85719, USA}
\email{david.sprayberry@noirlab.edu}

\author[]{G.~Tarl\'{e}}
\affiliation{University of Michigan, 500 S. State Street, Ann Arbor, MI 48109, USA}
\email{gtarle@umich.edu}

\author[]{B.~A.~Weaver}
\affiliation{NSF NOIRLab, 950 N. Cherry Ave., Tucson, AZ 85719, USA}
\email{benjamin.weaver@noirlab.edu}

\author[]{H.~Zou}
\affiliation{National Astronomical Observatories, Chinese Academy of Sciences, A20 Datun Road, Chaoyang District, Beijing, 100101, P.~R.~China}
\email{zouhu@nao.cas.cn}

%
\begin{abstract}
Environmental effects are a primary driver of elevated gas-phase
metallicities in galaxies around massive clusters, but the underlying
physical mechanisms for this satellite metallicity enhancement~(SME) are
still unclear. Using the Dark Energy Spectroscopic Instrument~(DESI) Data
Release 1, we present the first measurement of the average SME as a
function of projected cluster-centric distance. The resulting profile
reveals three distinct regimes: a steep decline from the cluster center, a
plateau near the cluster boundary, and an extended downturn across several
cluster radii. Remarkably, the complex shape and amplitude of this observed
SME profile are successfully reproduced in the EAGLE cosmological
simulation. Drawing insights from EAGLE, we develop a novel satellite
chemical evolution model to decompose the observed SME into physical
contributions from suppressed star formation, stellar mass loss, and
enriched gas inflow. Our analysis shows that continuous accretion of
enriched intracluster medium dominates the SME plateau within the cluster
virial radius, while mass loss and quenching jointly drive the rapid
metallicity decline in the cluster core. Our method disentangles the
impacts of three environmental processes on galactic chemical enrichment in
the cosmic web, providing a powerful framework for understanding cluster
galaxy evolution with current and future spectroscopic surveys.
\end{abstract}



\section{Introduction}
\label{sec:intro}

Galaxies in the cluster environment exhibit higher metallicity in their interstellar medium~(ISM) compared to those in the field~\citep{Shields1991,Skillman1996,Mouhcine2007,Pasquali2012,Peng2014}. This satellite metallicity enhancement~(SME) provides a unique probe into the key environmental processes that expedite the dynamical and chemical evolution of galaxies~\citep{Boselli2006,Blanton2009,Peng2015}. In this paper, we measure the SME profile as a function of distance from massive clusters using galaxies from the Dark Energy Spectroscopic Instrument~\citep[DESI;][]{Levi2013,DESI2016a,DESI2016b} survey Data Release 1~\citep[DR1;][]{DESIY12025} and compare it to the prediction by the EAGLE simulation~\citep{Schaye2015, Crain2015}, in hopes of developing a physically-motivated model of chemical enrichment for galaxies at the nexus of the cosmic web.

The ISM metallicity is commonly defined as $\Zism\equiv12{+}\log(\mathrm{O}/\mathrm{H})$, where $\mathrm{O}/\mathrm{H}$ indicates the oxygen-to-hydrogen abundance ratio in the ISM. The observed $\Zism$ is primarily correlated with the stellar mass $\ms$ of galaxies, yielding the well-established ``mass-metallicity relation''~\citep[MZR;][]{Lequeux1979,Trimonti2004,Savaglio2005,Erb2006,Mannucci2009,Andrews2013, Sanders2021,Curti2023}. Additionally, the observed scatter of $\Zism$ about the median MZR is correlated with the star formation rate~($\sfr$), resulting in a $\Zism{-}\ms{-}\sfr$ relation that does not evolve at least up to $z{\sim}3$~\citep[a.k.a., fundamental metallicity relation or FMR;][]{Mannucci2010,Lopez2010,Topping2021}.

In traditional gas-regulator models, the FMR is interpreted as a consequence of instantaneous chemical equilibrium between gas accretion and consumption~\citep{Bouche2010, Dave2012, Lily2013}. Challenging this view, \citealt{Lin2023} (hereafter referred to as \defcitealias{Lin2023}{LZ23}\citetalias{Lin2023}) demonstrated that an FMR naturally emerges from the non-equilibrium chemical evolution of galaxies in both the EAGLE simulation and the Sloan Digital Sky Survey~\citep[SDSS;][]{York2000}. They developed a non-equilibrium chemical evolution model~(NE-CEM) that can explicitly reconstruct the average histories of star formation~(SF), mass-loading, and gas accretion as functions of time for field galaxies at any observed $\ms$ and $\sfr$. A key finding of \citetalias{Lin2023} is that although galaxies spent most of their history out of chemical equilibrium, the self-similarity of their evolutionary histories produces a consistent FMR up to $z{=}3$. Consequently, the ISM metallicity of a typical star-forming galaxy in the field environment can be quantitatively understood within the NE-CEM framework.

However, after the star-forming galaxy falls into a massive cluster, it will likely experience the so-called ``delayed-then-rapid'' quenching process~\citep{Wetzel2013}, sustaining as a star-forming satellite for $2{-}4\,\gyr$ before a rapid transition onto the red-sequence~\citep[due to e.g., an increased ram pressure stripping in the cluster core;][]{Maier2019}. During the ``delayed'' phase, this satellite galaxy will experience stellar mass loss from the strong tidal field~\citep{Taylor2004,Green2021} and suppressed SF due to strangulation~\citep{Larson1980,Peng2015}, while still accreting residual gas from the metal-rich intracluster medium~\citep[ICM;][]{Schindler2005,Gupta2018}. As a result, the satellite metallicities will systematically deviate from those of isolated field galaxies at the same {\it observed} $\ms$ and $\sfr$. In this work, we present the first measurement of the scale-dependent SME profile for cluster galaxies using the DESI DR1 data, establishing a benchmark to evaluate subgrid physics in cosmological hydrodynamic simulations like EAGLE.

In theory, the scale-dependence of the SME profile is shaped by the combined environmental effects of mass loss, SF suppression, and enriched inflow at different radii away from the massive clusters. In particular, mass loss primarily reduces $\ms$ without changing the metallicity in the central region — the quantity directly measured by fiber-fed spectrographs like DESI. Meanwhile, suppressed SF modifies the chemical enrichment history~(CEH) of a galaxy by modulating the metal production vs. wind ejection~\citep{Lin2023}, and an enriched gas inflow directly elevates the metallicity of the ISM compared to the field galaxies where the accreted gas is typically chemically pristine~\citep{Finlator2008,Brooks2009}. While prior observational studies have provided qualitative assessments of the individual impacts of those processes on the SME~\citep{Ellison2009,Petropoulou2011,Williamson2016,SotilloRamos2021}, a rigorous understanding of the SME profile as a function of cluster-centric radius is still lacking. In this paper, we will draw analytic insights from the EAGLE simulation and build a simple yet comprehensive CEM for the satellite galaxies in DESI. Based on the NE-CEM developed by \citetalias{Lin2023}, our novel method aims to explicitly reconstruct the satellite histories of mass-loss, SF, and gas accretion inside the cluster, in order to accurately disentangle their contributions to the observed SME profile.

This is the first in a series of three papers that will systematically explore the SME physics using DESI DR1. The methodology presented in this paper will serve as the foundation for interpreting our more comprehensive DESI SME measurements in papers II and III.

We organize this paper as follows. We measure the overall SME profiles from
DESI DR1 and EAGLE in \S\ref{sec:sme_measurement} and develop a satellite
NE-CEM in \S\ref{sec:cemofsat}. In \S\ref{sec:ctrl_sample}, we describe our
method for identifying the physical processes that contribute to SME. From
analyzing the EAGLE simulation data, we decompose the observed SME profile
into three distinct contributions in \S\ref{sec:decomposition}. We
summarize our results and look to the future in \S\ref{sec:conclusion}.
We adopt a flat $\mathrm{\Lambda CDM}$ cosmology with
$\Omega_m{=}0.315$ and $h{=}0.674$~\citep{Planck2020} to calculate the distances for the DESI
measurements.

\section{Satellite metallicity enhancement in DESI and EAGLE}
\label{sec:sme_measurement}

As a first-cut analysis in this paper, we measure the overall SME profile of all star-forming galaxies in DESI DR1 surrounding clusters with $\mh{>}10^{12}h^{-1}M_\odot$ from the DESI halo-based cluster catalog. We defer a more comprehensive suite of SME measurements to the upcoming paper II, which will reveal the complex dependence of SME profile on stellar mass, $\sfr$, and halo mass. We describe the data and measurement method in \S\ref{subsec:data_desi} and \ref{subsec:sme_profile}, but the impatient readers can skip these two subsections to \S\ref{subsec:sme_comp}, where we compare the measured SME profiles between DESI and EAGLE.

\subsection{Galaxy and Cluster Samples in DESI DR1}
\label{subsec:data_desi}

\subsubsection{The DESI Star-forming Galaxy Sample}
\label{subsubsec:desisfg}

DESI is a prominent Stage-IV dark energy survey conducted with the $4$-meter Mayall telescope at Kitt Peak~\citep{Levi2013,DESI2016a,DESI2016b,DESI2022,DESI2024a,DESI2024b}. The DESI instrument obtains simultaneous spectra of almost 5000 objects~\citep{Guy2023, Poppett2024, Schlafly2023}. Our analysis is based on the Bright Galaxy Survey~(BGS) component of DESI DR1~\citep{DESIY12025, Adame2025}. This component includes the spectra of $6{,}280{,}198$ objects from the BGS Main Survey and $385{,}837$ from the Survey Validation~\citep[SV;][]{Hahn2023}, covering a redshift range of $0{<}z{<}0.6$. The BGS Main Survey consists of the BGS Bright sample~($r{<}19.5$) and BGS Faint sample~($19.5{<}r{<}20.175$), reaching galaxies ${>}2~\mathrm{mag}$ fainter than the Sloan Digital Sky Survey~\citep[SDSS;][]{York2000}. The SV was designed to operate similarly to the Main Survey, but achieving a much higher completeness~(${>}99\%$ for BGS Bright) over an area covering $180\,\mathrm{deg}^2$. In this work, we combine the BGS Bright, Faint, and SV samples to maximize the signal-to-noise ratio~(SNR) of our overall SME measurement.

From the combined BGS sample, we select galaxies with reliable redshift measurements by applying a set of quality cuts. For each galaxy, we use only the best available spectrum with \texttt{ZCAT$\_$PRIMARY==True}, and require \texttt{SPECTYPE==GALAXY}, \texttt{ZWARN==0}, and the $\Delta \chi^2$ of the best-fitting redshift above 40~\citep{Hahn2023}. We further impose a maximum redshift of $z{=}0.487$ to ensure that the $[\text{\ion{N}{2}}]\,\lambda 6584$ lines of all galaxies remain fully within the DESI spectral coverage~(3600\AA{}{-}9800\AA{}). After these redshift cuts, our galaxy sample is reduced by $24.9\%$ to $5{,}007{,}186$ galaxies, including $4{,}462{,}800$ from BGS Bright, $254{,}594$ from BGS Faint, and $289{,}792$ from SV.

The galaxy properties~(e.g., stellar mass and SFR) are derived from the combined DESI photometry and spectroscopy using the \textsc{Fastspecfit}\footnote{\url{https://fastspecfit.readthedocs.io/en/latest/}}~\citep{2023ascl.soft08005M} pipeline. In this work, we use version v2.1 of the \textsc{Fastspecfit} catalog of DR1 galaxies, which has already been made public as part of DR1~\citep{DESIY12025}. For the stellar population synthesis modeling, \textsc{Fastspecfit} assumes a \citet{Chabrier2003} initial mass function~(IMF), a constant solar metallicity, and a non-parametric star formation history~(SFH) with bursts.


We adopt the flux measurements of emission lines~($\mathrm{H}\alpha$, $\mathrm{H}\beta$, $[\text{\ion{O}{3}}]\,\lambda 5007$, $[\text{\ion{N}{2}}]\,\lambda 6584$). Regions with extreme reddening values~($\mathrm{E(B-V)}{>}0.8$) are excluded from this analysis. We then correct for the intrinsic dust extinction (i.e., due to the host galaxy) using the Balmer decrement, assuming case B recombination~($\mathrm{H}\alpha/\mathrm{H}\beta{=}2.87$) and the extinction law of~\citet{Cardelli1989}.

To select the star-forming galaxies, we adopt the criteria from~\citet{Kauffmann2003} based on the emission line ratios using the Baldwin–Phillips–Terlevich~\citep[BPT;][]{Baldwin1981} diagram. This BPT selection removes $21$ per cent active galactic nucleus~(AGN)-like galaxies from our sample. Following~\citet{Mannucci2010}, we further apply an SNR cut of $25$ on the $\mathrm{H}\alpha$ lines to select star-forming galaxies with robust overall line flux measurements~(with no additional SNR cuts on other emission lines). In the end, our selections result in $2{,}215{,}719$ galaxies in the star-forming galaxy sample before the metallicity measurements.

\subsubsection{Measurement of Gas-phase Metallicities}
\label{subsubsec:desimet}

We measure the gas-phase metallicity~$(\Zism)$ using the ``strong line'' method calibrated by~\citet[][hereafter referred to as M10]{Mannucci2010}. The M10 method makes use of the flux ratios of two sets of strong emission lines: $\mathrm{N}2\equiv[\text{\ion{N}{2}}]\,\lambda6584/\mathrm{H}\alpha$ and $\mathrm{R}23\equiv([\text{\ion{O}{2}}]\,\lambda3727{+}[\text{\ion{O}{3}}]\,\lambda4958,5007)/\mathrm{H}\beta$. In particular, the mapping between the line ratio and ISM metallicity is
\begin{equation}
    \label{eqn:strongline}
    \log R = a_0+a_1x+a_2x^2+a_3x^3,
\end{equation}
where $R$ is either $\mathrm{N}2$ or $\mathrm{R}23$, and $x{\equiv}\Zism{-}8.69$ is the ISM metallicity minus the solar value $Z_{\odot}{=}8.69$. The two sets of coefficients are $\{a_0,a_1,a_2,a_3 \}{=}\{-0.683,0.899,-0.523,-0.220\}$ and $\{ 0.718,-0.695,-0.622,-0.063\}$ for $\mathrm{N}2$ and $\mathrm{R}23$ diagnostics, respectively. Following M10, we only keep galaxies with both line ratios within their respective calibration ranges~($\log(\mathrm{N}2){<}0.35$ and $\log(\mathrm{R}23){<}0.90$) and the difference between two metallicity estimates within $0.25$ dex. This M10-specific selection reduces the size of the star-forming galaxy sample to $1{,}532{,}313$ galaxies with robust metallicity measurements. We adopt the average of the two metallicity estimates as our gas-phase metallicity $\Zism$ for each galaxy.

We do not apply any aperture corrections to the metallicity measurements. Since the emission line fluxes are measured from within the DESI fiber radius~($1.5\,\mathrm{arcsec}$), which corresponds to a comoving radius ${\sim}4\,\mathrm{kpc}/h$ at $z{=}0.2$, we are effectively measuring the metallicities in the central regions of galaxies. Meanwhile, the stellar masses derived from \textsc{Fastspecfit} correspond to the total stellar mass of each galaxy. Consequently, any mass loss in the outer regions does not affect our measured metallicities. We have also measured the overall SME profiles using another strong line-based method calibrated by~\citet{Curti2020} and the electron temperature-based direct method~\citet{Andrews2013}. The three measured SME profiles are consistent with one another, and the detailed comparison between the three sets of measurements will be presented in paper II.

\subsubsection{The DESI Cluster Sample}
\label{subsubsec:desienv}

To characterize the cluster environment, we employ the~\citet[][hereafter Y21]{Yang2021} group catalog derived by applying an adaptive halo-based group finder to the DESI Legacy Imaging Surveys DR9~\citep{Dey2019}. The Y21 group finder is primarily based on the photometric redshifts of galaxies, but has incorporated all DESI spectroscopic redshifts up to DESI DR2. The halo masses~($\mh$) in the catalog are obtained via abundance matching, derived from the rank-order of the total galaxy luminosities. The corresponding halo radii~($r_{180\mm}$) assume a spherical overdensity-based halo definition so that the average halo density within $r_{180\mm}$ is $180$ times the mean density of the Universe. To probe the environmental effects across a wide halo mass range, we select all the groups with $\mh{>}10^{12} h^{-1}M_\odot$ and richness ${>}3$ for our SME measurement. While the halo masses of most of the groups are below $10^{14} h^{-1}M_\odot$~(i.e., the conventional mass threshold for a cluster), we do not make any distinction between groups and clusters and use them interchangeably in this paper. We adopt the coordinates of the brightest cluster galaxies~(BCGs) as the cluster centers, and define the scaled cluster-centric radius as $\Rscale \equiv R_p / r_{180\mm}$, where $R_p$ is the projected cluster-centric distance.

The metallicity enhancement of cluster galaxies~(i.e., $\Rscale{<}1$) likely begins long before they first cross the cluster boundary~\citep{Zabludoff1998,Fujita2004}. To capture this ``chemical pre-processing'' effect~\citep{Gupta2018} in our SME measurements, we define an ``extended cluster environment'' using a cylindrical volume centered on each BCG with a projected radius of $10{\times}r_{180\mm}$~(i.e., $\Rscale{<}10$) and a line-of-sight velocity height of ${\pm}1500\,\mathrm{km/s}$. For simplicity, we apply the ``satellite'' label to all galaxies within this cylinder, although they mostly lie beyond the formal halo boundary. Meanwhile, we apply the ``field'' label to random galaxies drawn from our star-forming sample, without any selection on $\Rscale$.

\subsection{Measurement of the SME Profiles}
\label{subsec:sme_profile}

\begin{figure*}
    \centering
    \includegraphics[width=0.96\textwidth]{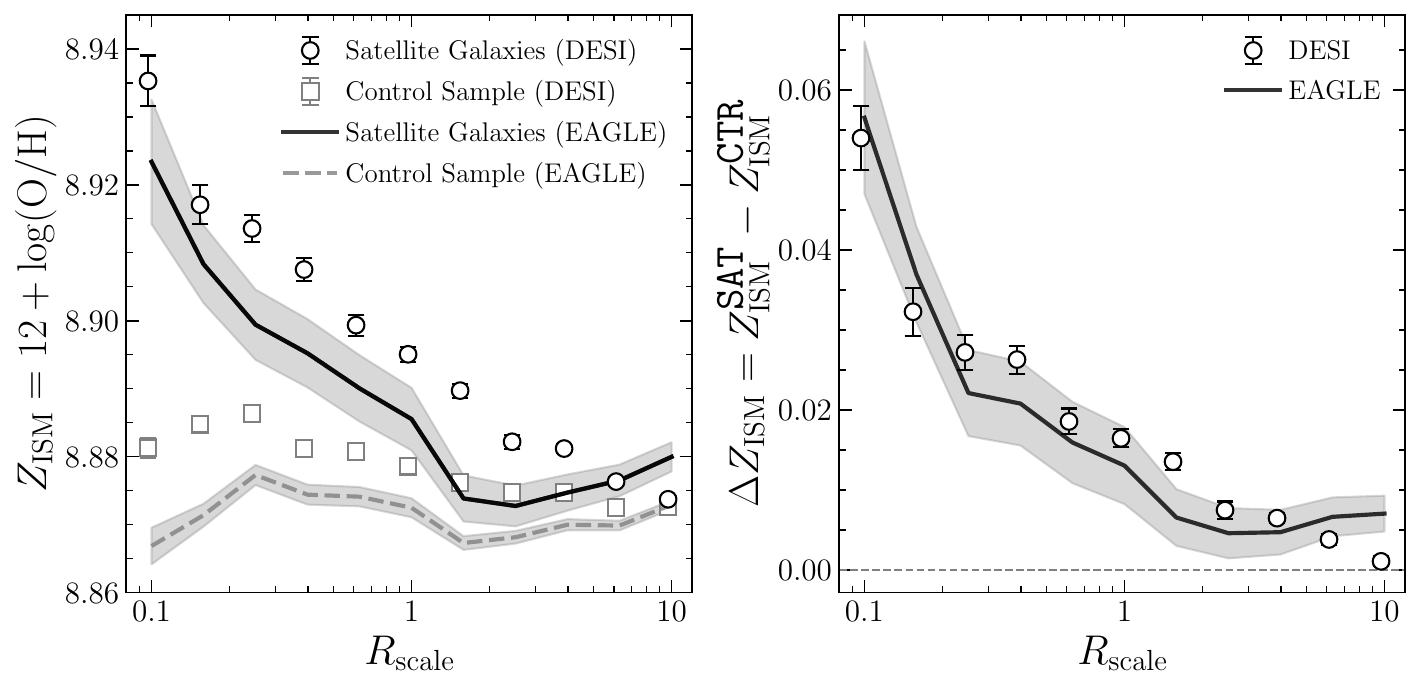}
    \caption{Comparison between the average gas-phase metallicity~(left)
    and SME~(right) profiles measured in DESI DR1 and the EAGLE simulation.
    Left: Average metallicity of the satellite~(black open circles with
    errorbars) and control~(gray squares with errorbars) samples as
    functions of projected radius $\Rscale$~(scaled by halo radius)
    measured in DESI DR1. Black solid~(gray dashed) curve with an
    uncertainty band is the average metallicity profiles of
    satellite~(control) galaxies in EAGLE. Right: The SME profiles derived
    from the left panel for DESI~(black open circles with errorbars) and
    EAGLE~(black solid curve with an uncertainty band).}
    \label{fig:SME_measurement}
\end{figure*}

We measure the overall SME profile $\Delta \Zism (\Rscale)$ from DESI as
\begin{equation}
    \label{eqn:sme}
    \Delta \Zism \left( \Rscale \right) = \Zism^{\sat}\left( \Rscale \right) - \Zism^{\texttt{CTR}}\left( \Rscale \right),
\end{equation}
where $\Zism^{\sat}$ and $\Zism^{\texttt{CTR}}$ are the average ISM metallicities of the satellite galaxies at $\Rscale$ and field galaxies with similar properties in the control sample, respectively. Therefore, the
key to robustly measuring SME profiles is to construct a proper control
sample at any given $\Rscale$.  Since the metallicity of a typical field
galaxy is primarily determined by its observed $\ms$ and $\sfr$~(by virtue
of the tight FMR), we construct a control sample of field galaxies for each
satellite sample at $\Rscale$ by matching their joint 3D probability
density distributions~(PDFs) of
$P^{\sat}(\ms,\sfr,z|\Rscale){=}P^{\texttt{CTR}}(\ms,\sfr,z|\Rscale)$. To
reduce the statistical noise in the $\Delta\Zism$ measurements, we require
that the size of the control sample to be at least ten times as larger as
that of the satellite sample at each $\Rscale$. Consequently, the $\Delta
\Zism$ measured from Equation~\ref{eqn:sme} can be safely interpreted as
the excess metallicity of satellite galaxies acquired due to various
physical processes in the extended cluster environment.

We use the EAGLE simulation to derive analytical insights into the physical processes governing the SME profiles observed by DESI. The EAGLE suite of cosmological simulations comprises a series of hydrodynamical simulations with different box sizes, particle numbers, and subgrid physics~\citep{Schaye2015, Crain2015}. In this work, we employ the ``Ref-L100N1504'' simulation, which is a box of $100\,\mathrm{Mpc}$~(comoving) on a side with $2{\times}1504^3$ particles~(equal number of dark matter and baryon particles). The particle mass is $1.81\times10^6\,h^{-1}\msun$ for gas/stars and $9.70\times10^6\,h^{-1}\msun$ for dark matter. This `reference' model has successfully reproduced several key scaling relations of galaxies, including the galaxy stellar mass function, stellar-to-halo mass relation, and MZR~\citep{Behroozi2013,Furlong2015,Schaye2015,DeRossi2017}. More important, \citetalias{Lin2023} used the ``Ref-L100N1504'' simulation to demonstrate that the NE-CEM model accurately describes the chemical enrichment of \emph{field} galaxies in EAGLE, providing a baseline framework for understanding the metallicity enhancement of EAGLE \emph{satellite} galaxies.

To measure the overall SME profile in EAGLE, we select all the
star-forming~($\sfr{>}0$) galaxies with $\ms{>}10^8\msun$ from the
$z{=}0.1$ snapshot of the simulation. Following \citetalias{Lin2023}, we
measure the ISM metallicities of EAGLE galaxies using
$\Zism\equiv12{+}\log(Z_\mathrm{O}/Z_\mathrm{H}/16)$, where $Z_\mathrm{O}$
and $Z_\mathrm{H}$ are the oxygen and hydrogen mass fractions,
respectively, in the ISM particles\footnote{This differs from
\citetalias{Lin2023}, where a constant $Z_\mathrm{H}{=}0.7$ was adopted to
simplify the analytic model.}. Halos in the EAGLE simulation are identified
as Friends-of-Friends (FoF) groups with $r_{200\mm}$~(roughly $5\%$ smaller
than $r_{180\mm}$) as the halo radii. We select the FOF groups with
$\mh{>}10^{12}\msun$ as our EAGLE cluster sample and adopt the same scheme
of constructing control samples when measuring $\Delta\Zism$ around these
EAGLE clusters. To facilitate the SME comparison between DESI and EAGLE, we
first measure the SME profile $\Delta \Zism(\Rscale)$ as a function of
projected distance $\Rscale{\equiv}R/r_{200\mathrm{m}}$ in EAGLE, but
switch to the isotropic SME profile $\Delta \Zism(\rscale)$ as a function
of the 3D scaled-distance~($\rscale{\equiv}r/r_{200\mathrm{m}}$) when
building the satellite NE-CEM model in the next Section.

\subsection{SME comparison between DESI and EAGLE}
\label{subsec:sme_comp}

We compare the measurements of SME profiles between DESI and EAGLE in Figure~\ref{fig:SME_measurement}. In the left panel, black circles and gray squares with errorbars are the metallicity profiles of the DESI satellite and control samples, respectively. All errorbar are computed from Jackknife resampling in both DESI and EAGLE. The metallicity profile of the control sample does not stay flat with $\Rscale$, because $P^\sat(\ms,\sfr,z|\Rscale)$ varies from one $\Rscale$ bin to another. For the same reason, the declining trend of the satellite metallicity profile is partly induced by the spatial variation in the composition of the satellite population. Black solid and gray dashed curves with uncertainty bands are the corresponding measurements from EAGLE.

The corresponding SME profiles are thus measured as the metallicity difference between the satellite and control samples, shown in the right panel of Figure~\ref{fig:SME_measurement}. Although both the amplitude and shape of the two EAGLE metallicity profiles are different from the DESI measurements in the left panel, the SME profile $\Delta \Zism(\Rscale)$ predicted by EAGLE~(solid circles) is remarkably similar to that measured from DESI galaxies~(open circles). In particular, both $\Delta \Zism$ profiles exhibit three distinct regimes of metallicity enhancement: a steep fall-off with increasing radius at $\Rscale{<}0.3$, a slow decline across halo boundary between $\Rscale{=}0.3$ and $\Rscale{=}2$, and a weak but non-zero excess beyond $\Rscale{=}2$. On the very large scales around $\Rscale{\sim}10$, the DESI profile steadily declines to zero; the EAGLE profile exhibits a plateau that stays positive, though the Jackknife errors in EAGLE are likely underestimated on those scales due to cosmic variance.

The strong agreement between the DESI and EAGLE SME profiles is highly
nontrivial. This suggests the subgrid physics model in EAGLE successfully
reproduces the \emph{excess} enrichment experienced by DESI galaxies in
cluster environments. To identify the dominant mechanisms responsible for
this agreement, we decompose the EAGLE SME profile at $z{=}0.1$ into
contributions from mass-loss, suppressed SF, and enriched inflow. In
addition, we expect the three SME regimes exhibited in the 2D projected
space~($\Rscale$) to be more sharply defined in the 3D real space~(i.e.,
using the 3D scaled cluster-centric distance $\rscale$). Therefore, we will
focus exclusively on the EAGLE simulation and switch to the 3D isotropic
SME profiles $\Delta \Zism(\rscale)$ for the rest of the paper until Figure~\ref{fig:sme3t2}, where we project the 3D decomposition into 2D to reproduce the DESI SME observation.

\section{Non-Equilibrium Chemical Evolution Model for Satellite Galaxies}
\label{sec:cemofsat}

\subsection{Components of a Satellite NE-CEM}
\label{subsec:satnecem}

\citetalias{Lin2023} demonstrated that the NE-CEM accurately describes the CEH of field galaxies in the EAGLE simulation and successfully reproduces the FMR in both EAGLE and SDSS. In this section, we adapt the NE-CEM framework to build a satellite NE-CEM by adding physics specific to the cluster environment: suppressed SF, and mass loss, and enriched inflow.

Among the three processes, suppressed SF is the most straightforward to model. We have verified that both the field and satellite galaxies in EAGLE follow the four parameter ``power-law exponential''~(hereafter shortened as \powexp) form of SFH,
\begin{equation}
    \label{eqn:powexp}
    \dot{M}_*(t) = \dot{M}_{*,0}\left(\frac{t-t_0}{\tausfh}\right)^{\kappa}
\exp \left(-\frac{t-t_0}{\tausfh}\right),
\end{equation}
where $t_0$ represents the formation time of the galaxy. This functional form exhibits a power-law rise with slope $\kappa$ followed by an exponential decline with a characteristic timescale $\tausfh$. As \citetalias{Lin2023} noted, Equation~\ref{eqn:powexp} has only two free parameters once the total formed mass and SFR at the observed epoch are specified. This constraint implies that the average SFH of satellite galaxies likely has a shallower $\kappa$ and a shorter $\tausfh$ than their field counterparts, resulting in suppressed SF at the later epochs. Since an accurate description of SF suppression is required for a clean SME decomposition, we directly adopt the best-fitting \powexp{} SFHs for satellite galaxies derived from EAGLE, deferring an analytic satellite SFH model to Paper III.

Regarding mass loss in satellites at a given $\rscale$, the dominant
physical mechanism is tidal stripping, which generally only incurs  mass
loss beyond the tidal radius $r_t$. Consequently, tidal mass loss does not
affect our metallicities observed in the central regions by DESI
fibers\footnote{Mass loss in the galaxy outskirts directly affects the
aperture-integrated metallicities due to non-zero metallicity
gradients~\citep{Zaritsky1994, Wang2019}.}. However, it reduces the total
stellar mass of the system, thereby indirectly elevating the satellite
metallicities relative to their field counterparts. For the sake of
accuracy, we directly measure the average mass loss for each satellite
sample at $\rscale$ in this work, deferring an analytic mass-loss model to
Paper III.

For incorporating enriched inflows in the satellite NE-CEM, we update the
original NE-CEM to allow accreted gas to have arbitrary metallicity.
\citetalias{Lin2023} assumed that accreted gas is chemically pristine.
While this is likely adequate for field galaxies, it requires modification
for satellite galaxies, which can actively accrete from the enriched ICM. A
large inflow of pristine gas can significantly dilute the ISM metallicity,
but in satellites, the accreted gas is already enriched. To model this
effect, we allow the satellite NE-CEM to switch, for instance, from a
pristine inflow before infall to an enriched one afterwards. Therefore, the
duration of enriched inflow for a satellite is equivalent to its time spent
in the cluster since infall, $\tauin$. We expect this infall time
$\tauin(\rscale)$ to be primarily determined by dynamical friction.

Below we describe our satellite NE-CEM framework that incorporates enriched inflow.

\subsection{Updated NE-CEM with enriched inflow}
\label{subsec:necem_enrichinflow}

To model the amount of metals brought in the ISM by enriched inflow, we need an explicit prescription for the gas accretion rate $\dmginf$. However, $\dmginf$ in NE-CEM is not modeled directly but implicitly through the time evolution of the gas reservoir. In particular, NE-CEM ties the mass of the star-forming gas $\mgas$ at any epoch $t$ to $\sfr$ via a volumetric Schmidt law~\citep{Kennicutt1998}
\begin{equation}
    \label{eqn:mgas}
    M_{\mathrm{{gas}}}= M_{\mathrm{{g,0}}}  \dot{M}_*^\epsilon,
\end{equation}
where the best-fitting parameters for EAGLE are $\log\,(M_{\mathrm{{g,0}}}/M_\odot){=}9.25{\pm}0.07$ and $\epsilon{=}0.93{\pm}0.05$. The time derivative of the gas mass is governed by star formation, galactic outflows, and gas accretion,
\begin{equation}
    \label{eqn:dmgas}
    \dot{M}_{\mathrm{{gas}}}= -(1-\frec)\dot{M}_*-\dmgout+\dmginf,
\end{equation}
where $\frec$ is the IMF-averaged recycle fraction, defined as the fraction of mass formed into stars that is returned to the ISM by supernovae and evolved stars, $\dmgout$ and $\dmginf$ represent the outflowing and inflowing rates of gas, respectively. To characterize the strength of outflows, we define the dimensionless mass-loading factor as $\eta{\equiv}\dmgout/\sfr$. Rearranging the preceding equation with this definition provides an expression for the gas accretion rate,
\begin{equation}
\label{eqn:dminf}
    \dmginf=\dot{M}_{\mathrm{gas}}+(1-\frec)\dms+\eta\dms.
\end{equation}

\citetalias{Lin2023} used a fitting formula~(their Eqn. 20) to describe the
dependence of mass-loading $\eta$ on $\ms$ and specific
$\sfr$~($\ssfr$) for the outflows
in regular star-forming galaxies. However, our satellite samples include
many star-forming but ``almost-quenched''~(i.e., with
$\ssfr{<}10^{-11}\,\mathrm{yr}^{-1}$) galaxies that do not follow the
\citetalias{Lin2023} fitting formula~(at least in EAGLE). To better
describe outflows in those galaxies, we update the best-fitting formula of
mass-loading from \citetalias{Lin2023} as
\begin{equation}
\label{eqn:mass_loading}
\log\!\left(\frac{\eta}{\eta_0}\right)
= f\,\!\left(\frac{M_*}{M_{*,0}}\right)^{\alpha}
 \left(\frac{\ssfr}{\ssfr_0}\right)^{\tilde{\beta}}
\end{equation}
where we set $f{=}0.49$, $\log\eta_0{=}{-}0.2$, $\log M_{*,0}{=}9.5$, and $\log\ssfr_0{=}{-}9.5$. For galaxies with $\ssfr{\ge}10^{-11}\,\mathrm{yr}^{-1}$, we adopt $\alpha{=}{-}0.15{\pm}0.02$ and $\tilde{\beta}{=}0.29{\pm}0.08$, and Equation~\ref{eqn:mass_loading} is thus equivalent with the Eqn. $20$ in \citetalias{Lin2023}; for the ``almost-quenched'' galaxies, we find that $\tilde{\beta}=0.5{\times}\log(\ssfr/10^{-11}\mathrm{yr^{-1}}){+}0.29$ provides a significantly better fit to the Eqn. $20$ in \citetalias{Lin2023}.

Meanwhile, the time evolution of oxygen mass in the ISM~($\mo$) is
\begin{equation}
    \label{eqn:modot}
    \modot= \mocc \dms - (1-\frec)\zo \dms -\eta \zo\dms + \zoinf\dmginf,
\end{equation}
where $\mocc$ is the IMF-averaged oxygen yield, defined as the mass of
oxygen produced by core-collapse supernovae~(CCSNe) and returned to the ISM
per solar mass of star formation, and $Z_\mathrm{O}$ and $\zoinf$ are the
oxygen mass fractions in the ISM and inflowing gas, respectively. On the
right-hand side~(RHS) of Equation~\ref{eqn:modot}, the four terms represent
the oxygen injected by CCSNe, incorporated into stars, ejected through
galactic winds, and accreted via gas inflow, respectively. Note that
\citetalias{Lin2023} adopted $\zoinf{=}0$ for field galaxies. For modeling
enriched inflows, we define the ``oxygen fraction ratio''~(OFR) between
inflowing gas and ISM as
\begin{equation}
    \label{eqn:finflow}
    \finf(t)= \frac{\zoinf(t)}{\zo(t)},
\end{equation}
and expect $0{<}\finf{<}1$ in the cluster environment.

In order to track the CEH of satellite galaxies, the key equation is the time derivative of $\zo$,
\begin{equation}
    \frac{\dd \zo}{\dd t}= \frac{\modot}{\mgas} -
    \frac{\dot{M}_{\mathrm{gas}}}{\mgas}\zo.
    \label{eqn:dzodtsimple}
\end{equation}
Plugging Equation~\ref{eqn:dminf},~\ref{eqn:modot} and~\ref{eqn:finflow} into Equation~\ref{eqn:dzodtsimple}, we arrive at
\begin{equation}
    \label{eqn:dzodt}
    \frac{\dd \zo}{\dd t}
    =\frac{\mocc}{\tau_*}
        - (1-\finf)\zo\left(\frac{1}{\tau_{\mathrm{dep}}} +\frac{\ddot{M}_*}{\dms} + \frac{\dot{\tau}_*}{\tau_*}\right).
\end{equation}
where $\tau_*{\equiv}\mgas/\dms$ is the gas consumption timescale and $\tau_{\mathrm{dep}}{\equiv}\tau_*/(1+\eta-r)$ is the gas depletion timescale. For constant $\tau_*$ and $\eta$, the ISM oxygen abundance would asymptotically approach an equilibrium value:
\begin{equation}
    \label{eqn:zoeq}
    \zoeq = \frac{\mocc}{(1-\finf)} \frac{\tilde{\tau}}{\tau_*},
\end{equation}
where $\tilde{\tau}$ is the ``harmonic difference timescale''
\begin{equation}
    \label{eqn:ttilde}
    \tilde{\tau}\equiv\frac{1}{\tau^{-1}_{\mathrm{dep}}-\tausfh^{-1}},
\end{equation}
introduced by~\citet{Weinberg2017}. Therefore, an enriched inflow boosts
the equilibrium oxygen fraction of the ISM by $1/(1{-}\finf)$ compared to a
pristine one.

This updated NE-CEM with arbitrary $\fin$ provides us a generic framework
for tracking the metallicity evolution of both satellite and field
galaxies. In particular, for a given satellite sample at $\rscale$, we
solve for its average CEH using Equation~\ref{eqn:dzodt} assuming a
time-varying $\fin$. For the metallicity evolution of field galaxies, we
adopt the original NE-CEM from \citetalias{Lin2023}, which is equivalent to
solving for Equation~\ref{eqn:dzodt} assuming a constant $\fin$ without
mass loss.

\section{Mathematical Decomposition of the SME}
\label{sec:ctrl_sample}

To ensure a \emph{clean} SME decomposition, we first design a parallel set of controlled experiments to unambiguously reveal the SME contribution from one \emph{single} physical process at a time~(i.e., a mathematical decomposition). Next, we use our satellite NE-CEM to clarify the underlying physics behind the mathematical decomposition, presenting a robust physical decomposition of the SME profile.

\subsection{Field Galaxies with Properties Matched to Satellite Galaxies}
\label{subsec:match_field}

\begin{figure}
    \centering
    \includegraphics[width=0.48\textwidth]{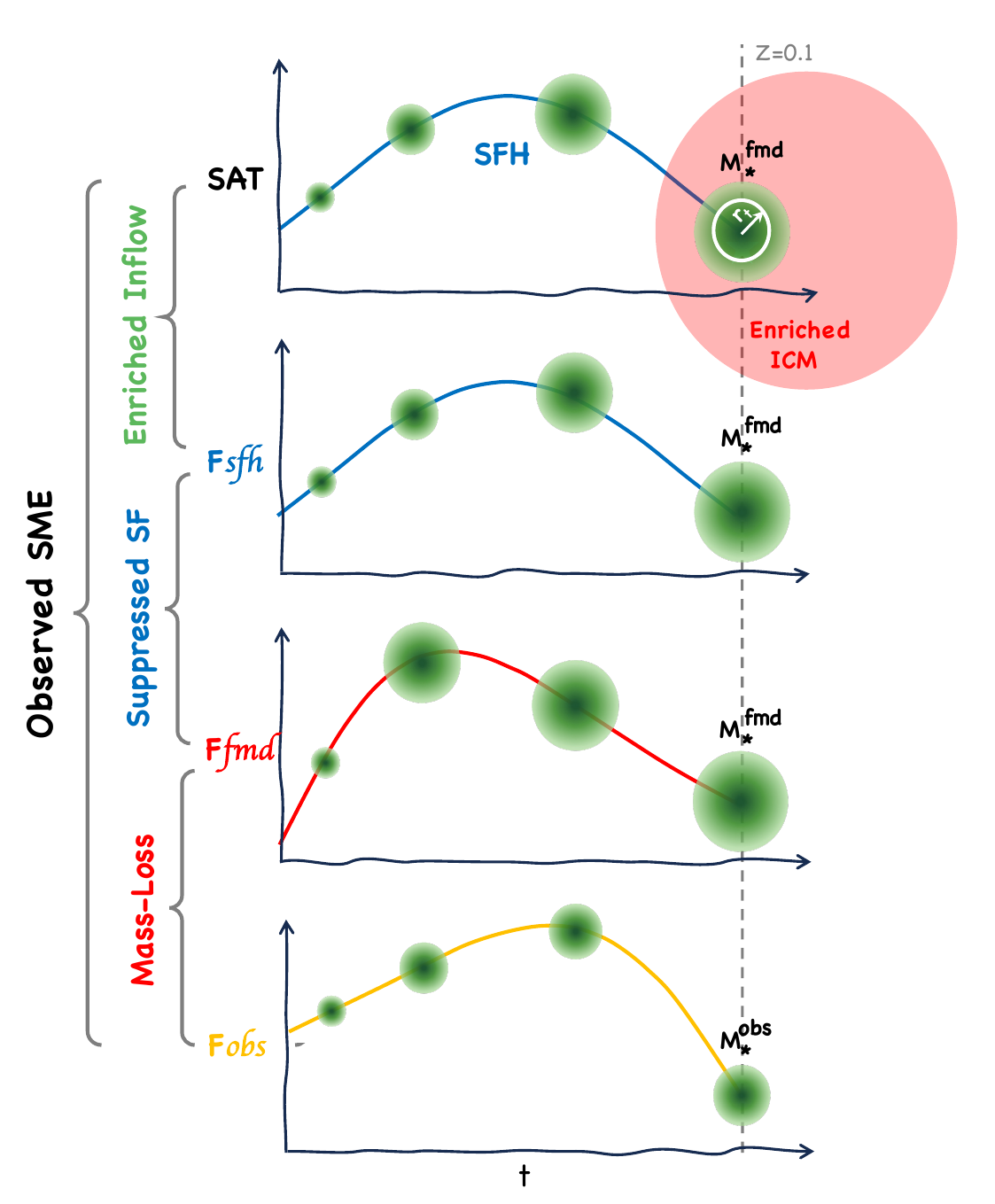}
    \caption{A cartoon version of Equation~\ref{eqn:decompose}, decomposing
    the observed SME into contributions due to three different physical
    processes~(annotated on the left). Each panel illustrates the typical
    evolutionary trajectory of a galaxy from one of the four samples in
    Table~\ref{tab:galaxy_sample}~(from top to bottom: \sat{}, \fthree{},
    \ftwo{}, \fone{}). The SFH is traced by the colored curve, through
    which the four green disks denote the stellar mass growth, with the
    disk size proportional to the formed mass $\msfmd(t)$. In the top
    panel, a \sat{} galaxy enters into the cluster at the final
    epoch~($z{=}0.1$), with its disk truncated at the tidal radius
    $r_t$~(solid white circle) and embedded in the enriched ICM~(red
    sphere). The stellar mass loss is thus indicated by the difference
    between $\msfmd$ and $\msobs$~(mass enclosed within $r_t$). The
    \fthree, \ftwo, and \fone{} are field galaxies matched to the \sat{} by
    SFH, $\msfmd$, and $\msobs$, respectively.}
    \label{fig:illustration_ctrl}
\end{figure}

In this subsection, we set the stage for the controlled experiments by defining three reference samples from the field galaxies in the simulation.

We start by measuring a set of physical quantities associated with star-forming satellite galaxies in the simulation. In particular, for a satellite galaxy observed at $z{=}0.1$, we follow the practice in \citetalias{Lin2023} and measure the in-situ SFH~($\dot{M}_*(t)$) by summing the SFRs of all its progenitors at the same epoch $t$, instead of considering only the main branch. The SFR observed at $z{=}0.1$ is thus
\begin{equation}
    \label{eqn:sfrobs}
    \sfrobs\equiv \dot{M}_*(t=t_{0.1}),
\end{equation}
where $t_{0.1}{=}12.45\,\gyr$ is the age of the Universe at $z{=}0.1$. From this SFH, we compute a \emph{formed} stellar mass as a function of time $t$ as
\begin{equation}
    \label{eqn:msform}
    \msfmd(t)=\int_0^t (1-\frec)\dot{M}_*(t')\,\dd t'.
\end{equation}

In an isolated field environment, the total mass of stars of a galaxy has ever formed by $t$, $\msfmd(t)$, is equal to the \emph{retained} mass at that epoch, $\msret(t)$, hence
\begin{equation}
    \label{eqn:msobs}
    \msret(t)=\msfmd(t) \quad \text{for field galaxies.}
\end{equation}
However, a satellite galaxy usually has $\msret(t)<\msfmd(t)$, and the difference between the two
\begin{equation}
    \label{eqn:mass_loss}
    \Delta\ms(t) = \msfmd(t) - \msret(t),
\end{equation}
quantifies the accumulated stellar mass-loss as a function of $t$, primarily due to tidal stripping in the cluster environment. For quantifying the physical state of each satellite galaxy with $\sfrobs$ at $z{=}0.1$, we define the following three {\it state variables}:
\begin{equation}
    \begin{aligned}[b]
    \msobs &\equiv \msret(t=t_{0.1}), \\
    \msfmd &\equiv \msfmd(t=t_{0.1}), \\
    \mathrm{SFH} &\equiv \dot{M}_*(t<t_{0.1}),
    \end{aligned}
    \label{eqn:prop_z01}
\end{equation}
where SFH is a vector variable measured from all the simulation snapshots before $z{=}0.1$. At any given $\rscale$, we thus characterize the satellite galaxy sample using $P(\msret,\msfmd,\mathrm{SFH})$, the joint distribution of $\msret$, $\msfmd$ and SFH of that sample. Accordingly, we define three types of reference samples using the field galaxies~(\fone{}, \ftwo{}, and \fthree{}), each with the PDF of one state variable matched to that of the satellite sample~(\sat{}). In addition, we ensure that all the four samples have the same $P(\sfrobs)$. The properties of the four samples are summarized in Table~\ref{tab:galaxy_sample}.

\begin{deluxetable}{ccr}
    \label{tab:galaxy_sample}
    \tablewidth{0pt}
    \tablecaption{Properties of four galaxy samples used for SME decomposition. Note that all four samples are additionally match in $\sfrobs$.}
    \tablehead{
    \colhead{Sample} & \colhead{Environment} & \colhead{Properties}}
    \startdata
    \sat & Cluster & $P(\msobs,\msfmd,\mathrm{SFH})$\\
    \fthree & Field & $P(\mathrm{SFH})$ matched to \sat\\
    \ftwo & Field & $P(\msfmd)$ matched to \sat\\
    \fone & Field & $P(\msobs)$ matched to \sat\\
    \enddata
\end{deluxetable}

Note that the \fone{} sample is equivalent to the \texttt{CTR} sample we constructed in \S\ref{subsec:sme_profile} when measuring the projected SME profiles in DESI. However, it is impossible to construct the \ftwo{} and \fthree{} samples in the observations, while using the EAGLE simulation allows us to construct all three reference samples from the simulated mock galaxies. As will be shown, our SME decomposition at each $\rscale$ depends critically on comparisons among different samples in Table~\ref{tab:galaxy_sample}.

\subsection{From Mathematical Decomposition to Physical Processes}
\label{subsec:method_decompose}

We measure the 3D isotropic SME profile from the EAGLE simulation as
\begin{equation}
    \label{eqn:dzism}
    \Delta \Zism (\rscale) = \Zism^{\sat}(\rscale) - \Zism^{\fone}(\rscale).
\end{equation}
At each $\rscale$, the RHS of the above equation can be mathematically separated into three terms as
\begin{equation}
    \label{eqn:decompose}
   \Delta Z_{\mathrm{ISM}} {=} \underbrace{(Z^{\sat} {-} Z^{\fthree})}_{\text{Enriched Inflow}} {+} \underbrace{(Z^{\fthree} {-} Z^{\ftwo})}_{\text{ Suppressed SF}} {+} \underbrace{(Z^{\ftwo} {-} Z^{\fone})}_{\text{Mass-Loss}},
\end{equation}
where the text below each term describes the physical process responsible for that term~(as will be explained further below). We omit the subscripts on the RHS of Equation~\ref{eqn:decompose} to avoid clutter.

Equation~\ref{eqn:decompose} provides the mathematical basis for our physically-motivated SME model. To illustrate this, Figure~\ref{fig:illustration_ctrl} uses schematic diagrams to demonstrate how each term in Equation~\ref{eqn:decompose} emerges from the three underlying physical processes. The four panels depict the evolutionary path of a typical galaxy from each of the four galaxy samples~(from top to bottom: \sat, \fthree, \ftwo, \fone). In each panel, the solid curve tracks the SFH of the galaxy on the $\sfr$ vs. $t$ diagram, arriving at the same $\sfrobs$ at $z{=}0.1$; the four green circular disks along the SFH curve represent the growth history of the galaxy through four epochs, with the disk sizes proportional to $\msfmd(t)$.

\begin{figure*}
    \centering
    \includegraphics[width=0.96\textwidth]{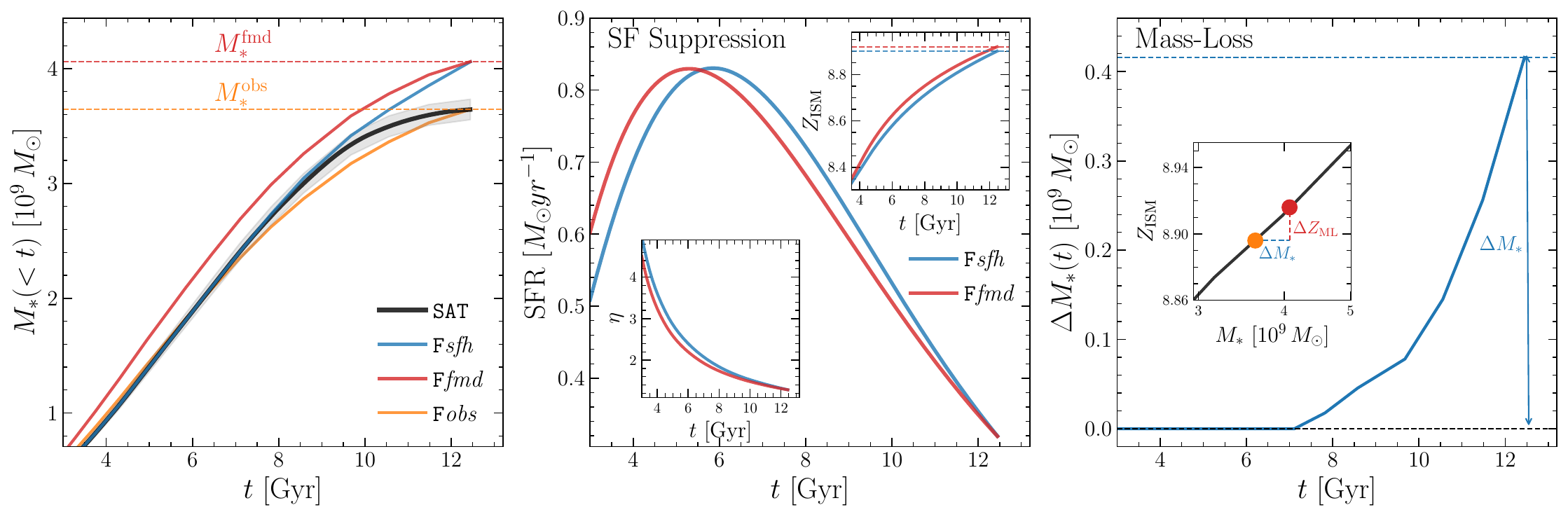}
    \caption{The average stellar mass growth~(left), star
    formation~(middle), and mass loss~(right) histories of relevant EAGLE
    galaxy samples at $\rscale{=}[0.2,0.4]$. Left: Colored curves indicate
    the stellar mass evolution of \sat{}~(black), \fthree{}~(blue),
    \ftwo{}~(red), and \fone{}~(orange) samples. Red and orange horizontal
    dashed lines denote the formed and observed stellar masses of the
    \sat{} sample at $z{=}0.1$, respectively. Middle: Red and blue curves
    are the average SFHs of \ftwo{} and \fthree{} samples, with the
    evolutions of their corresponding mass-loading factors and
    metallicities shown in the bottom-left and top-right inset panels,
    respectively. Right: Mass loss history of the \sat{} sample, computed
    from the difference between the blue and black curves in the left
    panel. The inset panel shows the MZR of galaxies with the same
    $\sfrobs$. Red and orange filled circles mark the location of the
    \ftwo{} and \fone{} galaxies, respectively.}
    \label{fig:galaxy_evolution}
\end{figure*}

In the top panel of Figure~\ref{fig:illustration_ctrl}, a \sat{} galaxy has recently falled into a cluster~(red sphere) and is thus embedded in the metal-enriched ICM at $z{=}0.1$. During the process, the \sat{} galaxy loses all the stellar mass beyond its tidal radius~($r_t$), denoted by the white circle within the green disk. Consequently, the mass enclosed within the white circle is the observed~(i.e., retained) mass $\msret$, whereas the total mass of the green disk is the formed mass $\msfmd$.

In the three bottom panels, each cartoon galaxy represents a field galaxy that has one state variable in Equation~\ref{eqn:prop_z01} exactly matched to the \sat{} galaxy, highlighting one particular physical process~(annotated vertically on the left) that contributes to the observed SME. We describe them as follows.
\begin{itemize}
    \item Enriched inflow~(second panel): an \fthree{} galaxy follows exactly the same SFH~(blue curve) as that of the \sat{} galaxy, but remained isolated in the field. Therefore, the main\footnote{The mass-loading histories are slightly different due to the stellar mass loss of the \sat{} galaxy.} difference in the metal enrichment between \sat{} and \fthree{} is whether the galaxy has accreted metal-enriched gas from the ICM in the last $\tauin\,\gyr$, contributing to the SME profile as the ``enriched inflow'' component~(i.e., $Z^{\sat}-Z^{\fthree}$).

    \item Suppressed SF~(third panel): an \ftwo{} galaxy has arrived at the same $\msfmd$ as that of \fthree{}, but follows the typical SFH~(red curve) of galaxies in the field rather than that of the \sat{} galaxy. The metallicity difference between \ftwo{} and \fthree{} is thus the SME contribution due to ``suppressed SF''~(i.e., $Z^{\fthree}-Z^{\ftwo}$).

    \item Mass-loss~(fourth panel): an \fone{} galaxy is a typical galaxy in the field with its stellar mass equal to $\msobs$~(i.e., the enclosed mass of the \sat{} galaxy within $r_t$). Since the \fone{} galaxy has a lower mass, it likely has a younger stellar age~(orange curve) than the \ftwo{} due to the so-called ``downsizing'' effect~\citep{Cowie1996}. Both the \ftwo{} and \fone{} galaxies are the typical field galaxies that land {\it perfectly} on the MZR, and the metallicity difference between the two can be entirely explained by their difference in $\ms$ — hence the SME contribution from ``mass-loss''~(i.e., $Z^{\ftwo}-Z^{\fone}$).
\end{itemize}

\section{Physical decomposition of the SME profile}
\label{sec:decomposition}

Having defined the three reference samples in \S\ref{subsec:match_field}, we now perform our physical decomposition of the SME profile using Equation~\ref{eqn:decompose}. As indicated in Table~\ref{tab:galaxy_sample}, these samples are primarily distinguished by their stellar mass growth histories. Accordingly, the left panel of Figure~\ref{fig:galaxy_evolution} of \sat{} galaxies at $\rscale{=}[0.2,0.4]$~(thick black curve) with that of the three reference samples: \fthree{}~(blue), \ftwo{}~(red), and \fone{}~(orange), echoing the schematic shown in Figure~\ref{fig:illustration_ctrl}. Note that all the curves are predicted by the best-fitting \powexp{} models, which provide accurate descriptions of the measurements from EAGLE~(not shown). The red and orange dashed horizontal lines mark $\msfmd$ and $\msobs$, respectively.

The red curve~(\ftwo), which represents the typical field galaxy observed with $\msfmd$, reaches the same \emph{formed} mass as the blue curve~(\fthree) but through different SFHs. Comparing these two would reveal the impact by suppressed SF. Meanwhile, the orange curve~(\fone), representing the typical field galaxy observed with $\msobs$, should arrive at a lower metallicity than that of $\ftwo$ according to the MZR. Thus, comparing these two directly along the MZR would reveal the mass loss term. By construction, the blue curve also represents the evolution of the formed stellar mass of the \sat{} sample. The \emph{retained}~(black) and \emph{formed}~(blue) stellar mass evolutions of the \sat{} sample start diverging at $t{\simeq}9\,\gyr$, likely the average epoch of infall for \sat{} galaxies at $\rscale{=}[0.2,0.4]$. The \sat{} galaxies would subsequently accrete gas from an enriched ICM. We examine the three comparisons in turn below, using the satellite sample at $\rscale{=}[0.2,0.4]$ as an example.

\subsection{Suppressed Star Formation and Mass Loss}
\label{subsec:SME_SFH_ML}

\begin{figure*}
    \centering
    \includegraphics[width=0.96\textwidth]{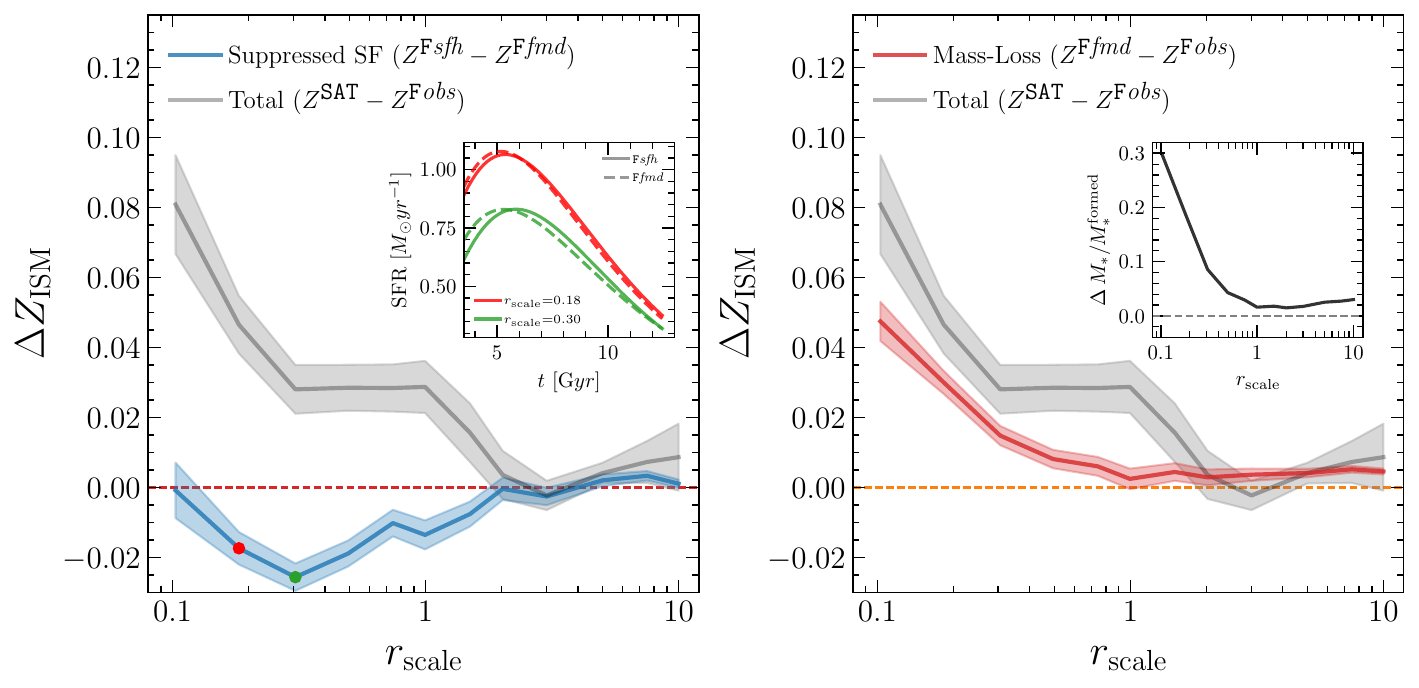}
    \caption{Contributions to the overall SME profile~(gray curve in each
    panel) due to suppressed SF (blue curve; left panel) and stellar mass
    loss (red curve; right panel), as functions of the 3D scaled radius
    $\rscale$ in EAGLE. The inset of the left panel shows the average SFHs
    of \fthree{}~(solid curves) and \ftwo{}~(dashed curves) samples at
    $\rscale{=}0.18$~(red) and $0.3$~(green), respectively. The inset of
    the right panel shows the fractional mass-loss as a function of
    $\rscale$ for the \sat{} galaxy sample.} \label{fig:decompose_SFH_ML}
\end{figure*}

We start our SME decomposition by examining the contribution caused by the SF suppression at $\rscale{=}[0.2,0.4]$. The middle panel of Figure~\ref{fig:galaxy_evolution} shows the \powexp{} SFHs of the $\msfmd$~(red curve; \ftwo{}) and SFH-matched~(blue curve; \fthree{}) galaxies. The two best-fitting SFH parameters are $\{\dot{M}_{*,0},t_0,\kappa,\tausfh \}=\{2.09,0.31,2.35,3.07 \}$ and $\{1.87,0.98,1.41,2.36 \}$ for \ftwo{} and \fthree{}, respectively.

Although the two SFHs produce exactly the same final stellar mass $\msfmd$ at $z{=}0.1$, the SFH of \fthree{} galaxies~(blue) follows that of the \sat{} galaxies, which exhibit a recent suppression after $t{=}9\,\gyr$, consistent with the reduced molecular hydrogen content found in satellite galaxies in the EAGLE simulation~\citep{Manuwal2023}. We predict the CEHs that correspond to the average SFHs of \fthree{} and \ftwo{} samples, as shown by the blue and red curves, respectively, in the top-right inset panel.

Interestingly, the SME contribution due to SF suppression is negative, i.e., $\Zism^{\fthree}-\Zism^{\ftwo}{<}0$~(comparing the two horizontal dashed lines in the top-right inset). This intriguing result can be readily understood using the NE-CEM framework. The \ftwo{} galaxies started forming stars much earlier, yielding a significantly more enriched ISM than the \fthree{} sample before $t{=}6\,\gyr$. The \fthree{} galaxies began to catch up in SF at $t{>}6\,\gyr$, but their average stellar mass remains lower than $\ftwo$, producing a higher mass-loading in the outflows~(blue curve in the bottom left inset panel). Therefore, the chemical enrichment of the \fthree{} sample remained fallen behind and fail to catch up with \ftwo{} by $z{=}0.1$.

\begin{figure}
    \centering
    \includegraphics[width=0.48\textwidth]{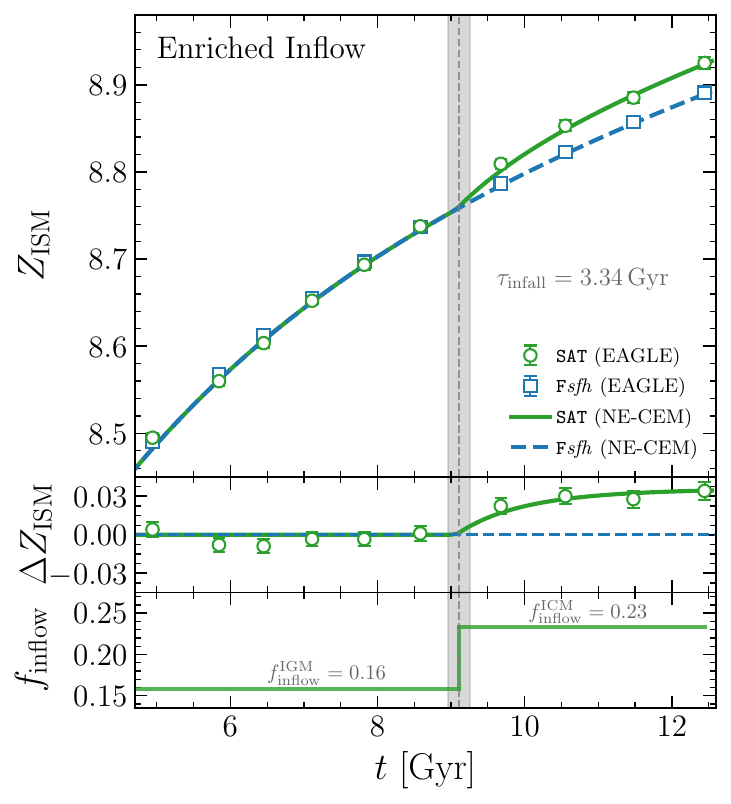}
    \caption{Satellite metallicity enhancement due to enriched flow at $\rscale{=}[0.2,0.4]$. Top: The average chemical enrichment histories of \sat{}~(green circles) and \fthree{}~(blue squares) galaxies measured from EAGLE. Green solid curve shows the prediction from our best-fitting satellite NE-CEM with $\tauin{=}3.34\,\gyr$ and $\ficm{=}0.23$, and blue dashed curve is the best-fitting prediction from a field NE-CEM with $\figm{=}0.16$. Middle: Metallicity difference between the \sat{} and \fthree{} samples as a function of time. Bottom: The time evolution of $\fin$ predicted by the best-fitting satellite NE-CEM model~(green curve).}
    \label{fig:enriched_ceh}
\end{figure}

The SME contribution due to mass-loss is illustrated in the right-most panel of Figure~\ref{fig:galaxy_evolution}. The blue curve is the average mass-loss experienced by the \sat{} galaxies as a function of time, computed from the difference between the black and blue curves in the left-most panel~(Equation~\ref{eqn:mass_loss}). The \sat{} galaxies start losing a significant amount of stellar mass upon infall~($t{\sim}9\,\gyr$), and the fractional mass-loss reaches ${\sim}10\%$ by $z{=}0.1$.

To convert this mass loss into a metallicity enhancement, we show the MZR~(black curve; $\Zism{=}0.40{\times}\log\ms{+}5.07$) of EAGLE galaxies at fixed $\sfrobs{=}{10^{-0.5}}\,\msun yr^{-1}$ in the inset panel. The red and orange circles indicate the locations of two typical galaxies from the \ftwo{} and \fone{} samples, respectively. Consequently, we directly read off the SME contribution due to mass-loss $\Delta\ms$ as $\Zism^{\ftwo}{-}\Zism^{\fone}{=}0.40{\times}\Delta\left(\log\ms\right){=}0.015$.

Following the decomposition method outlined in
Figure~\ref{fig:galaxy_evolution}, we derive the SME contributions from the
suppressed SF and mass-loss as functions of $\rscale$, shown in the left
and right panels of Figure~\ref{fig:decompose_SFH_ML}, respectively. In
both panels, the gray curve with a shaded uncertainty band indicate the
total SME profile of EAGLE satellite galaxies as a function of the 3D
scaled distance $\rscale$. As expected from \S\ref{sec:sme_measurement},
the three distinct regimes of SME we identified from the projected profile
become even more prominent in the 3D profile, exhibiting a steep inner
slope at $\rscale {<} 0.3$, a plateau at $0.3 {<} \rscale {<} 1$, and a
shallow decline at $1 {<} \rscale {<} 3$ before eventually approaching zero
at $\rscale{>} 3$.

In the left panel of Figure~{\ref{fig:decompose_SFH_ML}}, the blue curve with an error band indicates the SME contribution due to suppressed SF, derived from the metallicity difference between \fthree{} and \ftwo{} as a function of $\rscale$.

This SME component is negative on all scales below $\rscale{=}2$ and exhibits a non-monotonic behavior, with a minimum of $\Delta\Zism{=}{-}0.022$ at $\rscale{=}0.3$. This is consistent with the ``delayed-then-rapid'' quenching scenario, in which galaxies experience the maximum suppression of SF right before they become rapidly quenched around $\rscale{\sim}0.3$. Starting at the infall~($\rscale{\sim}1$), SFH-matched galaxies~(\fthree) show little difference in SFH compared with their $\msfmd$-matched counterparts~(\ftwo). As $\rscale$ decreases, the SFH of satellites becomes increasingly suppressed, resulting in a growing metallicity deficit. However, once the \sat{} galaxies reach the cluster core at $\rscale{\sim}0.3$, most have concluded the ``delayed'' phase of residual SF and begin to be rapidly quenched. The small number of star-forming galaxies found at $\rscale{<}0.3$ are likely systems with substantial central gas reservoirs, allowing their SFHs to remain comparable to those in the field.

To better understand the origin of this minimum, we compare the levels of SF suppression between $\rscale{=}0.3$~(green curves) and $\rscale{=}0.18$~(red curves) in the inset panel. For each $\rscale$, the SFH of the \fthree{}~(i.e., SFH-matched) and \ftwo{}~($\msfmd$-matched) samples are shown by the solid and dashed curves, respectively. The average formed stellar mass of the \sat{} galaxies at $\rscale{=}0.18$~($\log\msfmd{=}9.69$) is higher than that at $\rscale{=}0.3$~($\log\msfmd{=}9.58$), consistent with the shorter timescale of dynamical friction for the more massive galaxies. This explains the higher amplitude of the SFHs at $\rscale{=}0.18$ than at $\rscale{=}0.3$. More important, the level of SF suppression at $\rscale{=}0.3$~(compare the two green curves) is much stronger than that at $\rscale{=}0.18$~(compare the two red curves).

Switching to the right panel of Figure~\ref{fig:decompose_SFH_ML}, the red curve with an uncertainty band shows the SME contribution due to mass loss, computed from the metallicity difference between the \ftwo{}~($\msfmd$-matched) and \fone{}~($\msobs$-matched) samples. Unlike the suppressed SF, the mass-loss component is positive across all scales. Since the amplitude of this component is linearly proportional to the amount of mass loss at $\rscale$, the declining trend with increasing $\rscale$ directly reflects the dependence of fractional mass-loss on radius, shown by the black curve in the inset panel. The infalling galaxies often started losing mass due to an enhanced tidal field and frequent fly-bys on scales about several times the halo radius~\citep{Moore1996}, producing the positive plateau in the mass-loss component at $\rscale{=}1{-}10$.

Combining the two panels of Figure~\ref{fig:decompose_SFH_ML}, we find that the SF-suppression and mass-loss components have similar declining shapes with increasing radius at $\rscale{<}0.3$, consistent with the sharp drop-off exhibited by the total profile in the inner region. Beyond $\rscale{=}0.3$, the two components have opposite signs but comparable amplitudes, suggesting that the sum of the two cannot explain the plateau exhibited by the total SME profile at $0.3{<}\rscale {<}1$. We thus expect the third contribution, i.e., enriched inflow, to be the most dominant process that boosts the metallicity of satellites on those scales.

\subsection{Enriched Inflow}
\label{subsec:SME_EI}

In order to extract the SME component due to enriched
inflow, we adopt the satellite NE-CEM developed in \S\ref{sec:cemofsat} with $\fin(t)$ varing as a step-function
\begin{equation}
    \label{eqn:fin}
    \fin(t) =
    \begin{cases}
         0.16  & t\le t_{0.1} - \tauin \\
        \fin^{\mathrm{ICM}}    & t > t_{0.1} - \tauin,
    \end{cases}
\end{equation}
where
\begin{equation}
    \label{eqn:finicm}
    \log \fin^{\mathrm{ICM}} \equiv Z_{\mathrm{ICM}} - \Zism,
\end{equation}
is the metallicity difference between ICM and ISM, i.e., the logarithmic ICM-to-ISM OFR. We set the pre-infall value of $\fin$ to be $0.16$, estimated using all the star-forming field galaxies at $z{=}0.1$ in EAGLE. Consequently, our satellite NE-CEM has only two parameters, the infall time $\tauin$ and the ICM-to-ISM OFR $\fin^{\mathrm{ICM}}$, while the SFH and mass loss history are directly measured from the simulation.

\begin{figure*}
    \centering
    \includegraphics[width=0.96\textwidth]{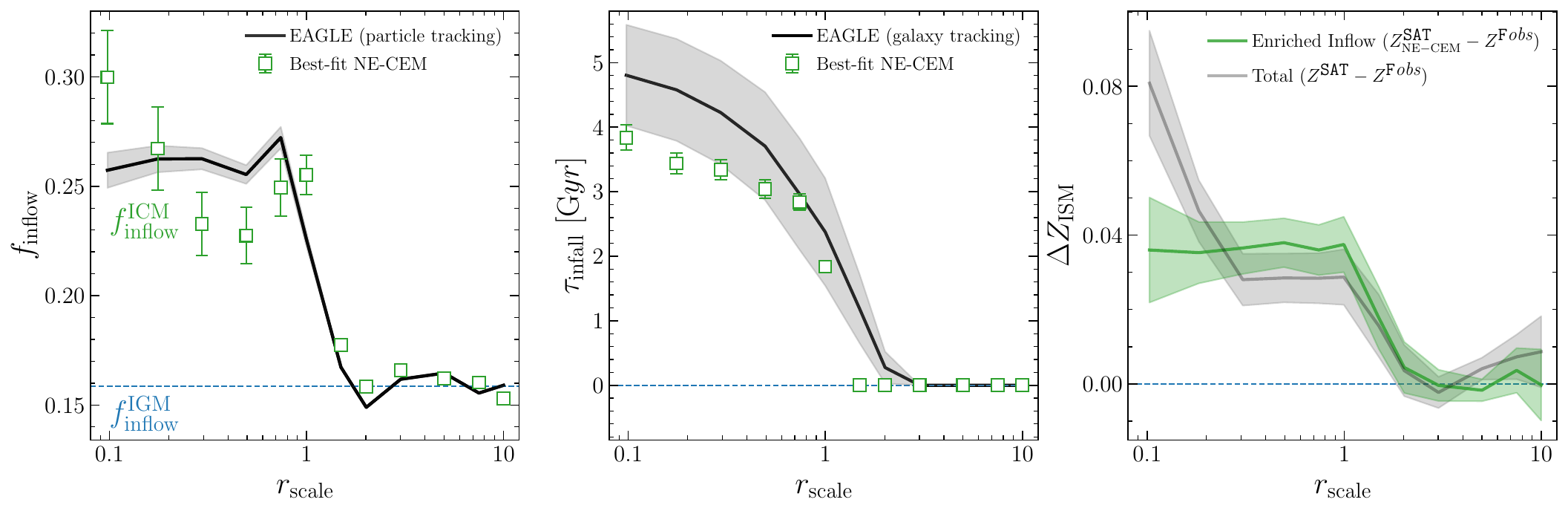}
    \caption{Contribution to the overall SME profile due to enriched inflow, as predicted by our satellite NE-CEM. Left: Oxygen fraction of the inflow~(relative to the ISM) $\fin$ as a function of $\rscale$. Black curve with a gray uncertainty band indicate the direct measurement from EAGLE, while squares with errorbars are the constraints from our satellite NE-CEM. Middle: Similar to the left panel, but for the infall time $\tauin$. Right: Green curve with an uncertainty band indicates the SME contribution due to enriched inflow, predicted by our best-fitting satellite NE-CEM. Gray curve is the overall SME profile measured from EAGLE.}
    \label{fig:decompose_EI}
\end{figure*}

Figure~\ref{fig:enriched_ceh} compares the CEHs of the \sat{}~(green circles) and \fthree{}~(blue squares) samples at $\rscale{=}[0.2,0.4]$, measured directly from the simulation. The blue dashed curve represents the prediction for \fthree{}~(i.e., SFH-matched) from our field NE-CEM~(i.e., setting $\figm{=}0.16$ in Equation~\ref{eqn:dzodt}), providing an excellent description of its CEH. Next, we fit the \sat{} CEH~(green circles) with our satellite NE-CEM using Equation~\ref{eqn:dzodt} and~\ref{eqn:fin}, while adopting the same \powexp{} SFH as that used for predicting the \fthree{} CEH. The chemically-inferred constraints on the infall timescale and ICM-to-ISM OFR are $\tauin{=}3.34{\pm}0.15\,\gyr$ and $\fin^{\mathrm{ICM}}{=}0.23{\pm}0.02$, respectively, with the best-fitting prediction shown as the green solid curve. The middle subpanel shows the metallicity difference between the \sat{} and \fthree{} samples as a function of time. Upon entering the cluster environment, satellite galaxies undergo rapid metallicity enhancement within the first gigayear, due to the sudden jump in $\finf$~(Equation~\ref{eqn:fin}; green curve in the bottom subpanel). After this initial increase, the metallicity difference relative to the field remains constant until $z{=}0.1$. Similar to the $\rscale{=}[0.2,0.4]$ bins shown in Figure~\ref{fig:enriched_ceh}, our best-fitting satellite NE-CEM provides excellent descriptions of the CEHs of \sat{} galaxies in all other $\rscale$ bins~(not shown), providing constraints on $\tauin$ and $\ficm$ as functions of $\rscale$.

\begin{figure*}
    \centering
    \includegraphics[width=0.96\textwidth]{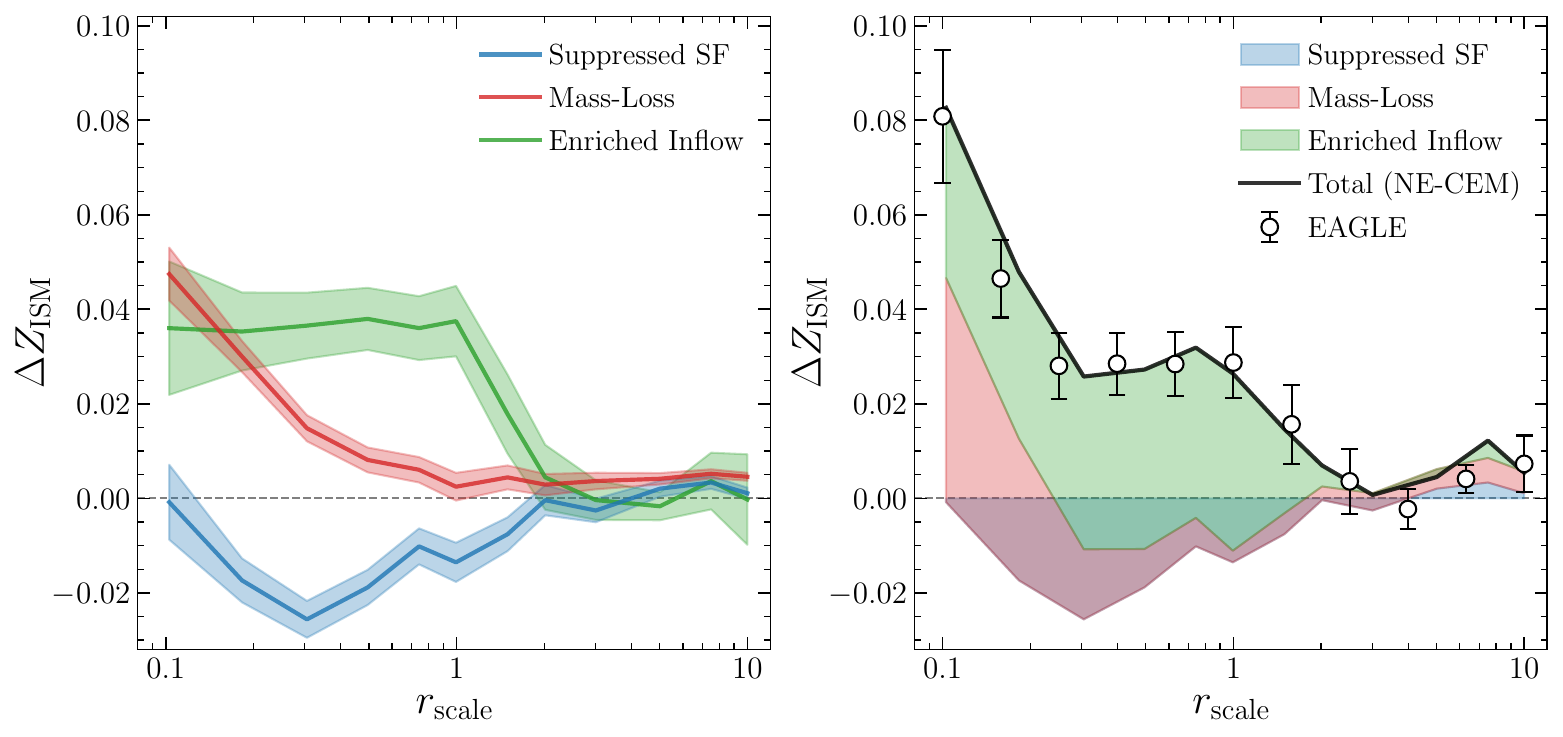}
    \caption{Final decomposition of the 3D overall SME profile in EAGLE. Left: Blue, red, and green bands indicate the SME components due to suppressed SF, mass-loss, and enriched inflow, respectively, as functions of $\rscale$. Right: Comparison between the overall SME profiles reconstructed~(black curve) by stacking the three contributions~(colored layers) and directly measured from EAGLE~(open circles with errorbars).}
    \label{fig:decomposition}
\end{figure*}

We now ask whether our chemically-inferred values of $\tauin$ and $\ficm$ are consistent with the direct measurements from the simulation. In the left panel of Figure~\ref{fig:decompose_EI}, we show our chemically-inferred $\ficm$ in different $\rscale$ bins as green squares with errorbars, while the directly-measured metallicity ratios between ICM and ISM gas particles are indicated by the black curve with a gray uncertainty band. The simulation measurement has a flat profile within $\rscale{=}0.5$, while the chemically-inferred profile shows a declining trend with increasing $\rscale$. The cause of this discrepancy is likely complex — the chemically-inferred $\fin$ is a somewhat CEH-averaged value over the course of the satellite evolution before reaching $\rscale$, while the simulation-measured value is instantaneous at that $\rscale$. Nevertheless, the two sets of $\fin(\rscale)$ profiles agree reasonably well over the entire range of $0 {<} \rscale {<} 10$, including both the overall amplitude of $\fin$ within $\rscale{=}1$ and the sharp transition across the halo boundary.

In the middle panel of Figure~\ref{fig:decompose_EI}, we compare our chemically-inferred $\tauin$ profile~(green squares with errorbars) with the direct measurement from simulation (black solid curve with an uncertainty band). We measure the average $\tauin$ of the \sat{} sample at different $\rscale$ by tracking the main branch of a galaxy’s merger tree and identifying the epoch at which that galaxy changed its identity from being a central galaxy~($\texttt{SubGroupNumber}{=}0$) to a satellite ($\texttt{SubGroupNumber}{>}0$). The measurement uncertainty includes both the error on the mean in the sample and the error associated with the interpolation between two snapshots. The simulation-measured $\tauin(\rscale)$ profile remains positive at $1 {<} \rscale {<} 2$, because the FoF clusters may have satellite beyond $r_{200\mathrm{m}}$. Overall, the chemically-inferred $\tauin$ is shorter than the simulation measurement, but the two are roughly consistent within $1\sigma$.

The overall agreement between the two types of measurements is encouraging, suggesting that we can potentially infer the dynamical history of satellite galaxies from their observed metallicity enhancement.

Repeating the analysis shown in Figure~\ref{fig:enriched_ceh}, we compute the SME contribution~($\Zism^{\sat}-\Zism^{\fthree{}}$) due to enriched inflow at different $\rscale$, shown as the green curve with an uncertainty band in the right panel of Figure~\ref{fig:decompose_EI}. Same as in Figure~\ref{fig:decompose_SFH_ML}, the gray curve with an uncertainty band represents the total SME profile in the EAGLE simulation. As we expected in \S\ref{subsec:SME_SFH_ML}, the SME component due to enriched inflow dominates the scale between $\rscale{=}0.3$ and $\rscale{=}1$, and stays flat with an amplitude of $\Delta\Zism{\sim}0.037$. Beyond $\rscale{=}1$, the enriched-inflow component declines rapidly to zero at $\rscale{\sim}2{-}3$.

The SME of $\Delta\Zism{\sim}0.037$ can be understood as follows. From Equation~\ref{eqn:zoeq}, we expect the boost in metallicity to be $(1{-}0.16)/(1{-}\fin^{\mathrm{ICM}}){=}1.09$ if the satellites have reached chemical equilibrium. This $9$ per cent boost is consistent with $10^{\Delta\Zism}{\sim}1.09$. The fact that the green curve is roughly flat within $\rscale{=}1$ indicates that the timescale for this metallicity boost due to ICM is short compared to $\tauin$, as can be seen from the middle panel of Figure~\ref{fig:enriched_ceh}. Therefore, this elevation in metallicity does not require equilibrium, as most of the satellites at $\rscale{\sim}1$ fell into clusters recently and have yet to adjust to chemical equilibrium.

\subsection{Comparison of the Three SME Components}

Putting all the pieces together, we summarize the results of our comprehensive SME decomposition in Figure~\ref{fig:decomposition}. In the left panel, blue, red, and green curves with error bands indicate the SME contributions from suppressed SF, mass-loss, and enriched
inflow, respectively. The three curves are derived separately earlier in the paper according to Equation~\ref{eqn:decompose}. We find that the three components have very different radial dependence within $\rscale {=} 2$, thereby shaping the three distinct regimes in the total SME profile.

The right panel of Figure~\ref{fig:decomposition} provides a more visually appealing way for understanding the shape of the total SME profile. The simulation measurement is shown by the open circles with errorbars. Underneath these data points, we present a stacked version of the SME decomposition, illustrating each component by the area covered by its respective color listed in the legend. The thick black curve indicates the sum of the three contributions, which provides an excellent description of the data points measured directly from the simulation.

The enriched-inflow component dominates the SME in the outer region of the cluster between $\rscale{=}0.3{-}1$; but in the inner core of clusters~($\rscale{<}0.3$), the mass-loss term starts to take over and rise towards $\rscale{=}0$. However, this SME increase due to mass-loss is largely offset by the negative contribution due to suppressed SF on scales $\rscale{=}0.3{-}1$. Interestingly, the suppressed-SF component has a V-shaped profile with a minimum at $\rscale{=}0.3$ , possibly driven by the ``delayed-then-rapid'' quenching of satellites. Thus, the sharp increase of both the mass-loss and suppressed-SF components towards $\rscale {=} 0$ produces an even steeper slope in the total SME profile at $\rscale {<} 0.3$.

\begin{figure*}
    \centering
    \includegraphics[width=0.96\textwidth]{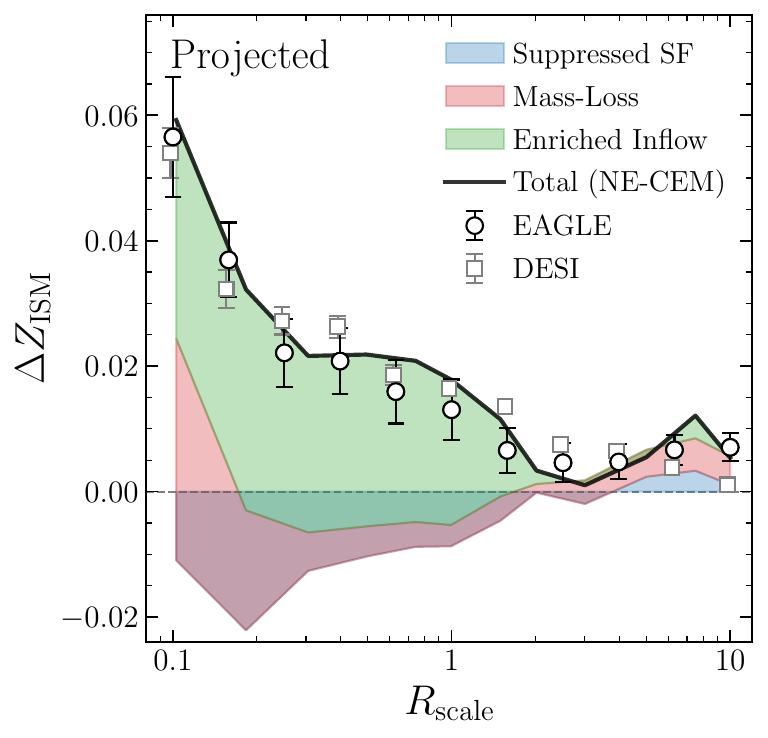}
    \caption{Same as the right panel of Figure~\ref{fig:decomposition}, but
    in 2D projected space. Black circles with errorbars are the EAGLE
    measurement in 2D, while gray squares with errorbars are the overall
    SME profile measured in DESI DR1. The combination of our
    physically-motivated decomposition and the satelite NE-CEM model
    provides an excellent description of the SME profile measured in DESI.}
    \label{fig:sme3t2}
\end{figure*}

Finally, the excellent agreement between the direct measurement and our SME
model prediction demonstrates that, our satellite NE-CEM is able to capture
the key physical processes that drive the SME phenomenon in the EAGLE
simulation. We are hopeful that our physically-motivated model will
describe, at least qualitatively, the satellites in the real Universe as
well, given the similarity of the SME profiles between DESI and EAGLE. To
better illustrate how our decomposition can explain the DESI SME profile
measured in the projected space, we convert our three SME components from
3D to 2D using
\begin{equation}
    \label{eqn:sme3t2}
    \Delta \Zism(\Rscale) =\frac{\int_{0}^{\hscale^{\mathrm{max}}}\Delta
    \Zism (\tilde{r})n_s(\tilde{r})\,\dd \hscale}{\int_{0}^{\hscale^{\mathrm{max}}}n_s(\tilde{r})\,\dd \hscale},
\end{equation}
where $\hscale$ is the line-of-sight distance normalized by halo radius,
$\hscale^{\mathrm{max}}$ is the integration limit which we set to $10$,
$n_s$ is the 3D satellite number density profile around clusters, and
$\tilde{r}{\equiv}\sqrt{\Rscale^2+\hscale^2}$. The results are shown in
Figure~\ref{fig:sme3t2}. Open circles and squares with errorbars are the projected SME profiles measured from EAGLE and DESI~(same as those
shown in Figure~\ref{fig:SME_measurement}), respectively. The blue, red,
and green stacks are the 2D projected contributions from suppressed SF,
mass-loss, and enriched inflow, respectively. The black solid curve is the
sum of the three projected components, providing a good description of the
DESI observation.

\section{Conclusion}
\label{sec:conclusion}

In this paper, we present the first measurement of the average SME profile as a function of projected distance~(scaled by halo radius) away from galaxy clusters, $\Delta \Zism(\Rscale)$, using the Bright Galaxy Survey galaxies from DESI DR1. The observed SME profile exhibits three distinct regimes: a steep decline with increasing radius at $\Rscale{<}0.3$, a gradual decrease across the halo boundary over $0.3{\le}\Rscale{\le}2$, and at $\Rscale{>}2$, a weak but non-zero enhancement extending to $\Rscale{\sim}10$.

Remarkably, we find that both the shape and amplitude of the DESI SME profile are well reproduced by the EAGLE hydrodynamical simulation at $z{=}0.1$. This agreement suggests that the subgrid physics implemented in EAGLE likely captures the essential environmental processes regulating chemical enrichment in satellites. Analyzing the EAGLE simulation, we identify three dominant mechanisms that drive the SME around massive clusters: the suppression of star formation experienced by satellite galaxies, the stellar mass loss due to the strong cluster tidal field, and the accretion of enriched gas from the ICM.

Accordingly, we develop a set of clean, controlled experiments to decompose the 3D isotropic EAGLE SME profile $\Delta\Zism(\rscale)$ into contributions from three physical processes: suppressed star formation, mass loss, and enriched inflow. We find that the suppressed star formation produces a negative SME contribution~($\langle \Delta \Zism\rangle {\sim}{-}0.02$) with a minimum of ${\simeq}{-}0.026$ at $\rscale{\simeq}0.3$, likely caused by the ``delay-then-rapid'' quenching process of satellite galaxies. Meanwhile, mass loss shifts satellite galaxies horizontally on the metallicity vs. stellar mass diagram to the left, thereby producing a~(pseudo-) positive SME effect~($\langle \Delta \Zism\rangle {\sim}0.02$). Having comparable magnitudes but different signs, the contributions from the suppressed star formation and mass loss roughly cancel each other at $0.3{<}\rscale{<}1$. Within the cluster core at $\rscale{\le}0.3$, however, the two conspire to steepen the increase of the SME profile towards the cluster center.

We find that the enriched inflow dominates the SME in the outer regions of clusters~($0.3{<}\rscale{<}1$), with a roughly $\rscale$-independent contribution of $\Delta \Zism{\simeq}0.037$. To better understand the metallicity evolution of satellite galaxies, we develop a satellite NE-CEM that extends the NE-CEM framework of \citetalias{Lin2023} to incorporate enriched inflows~(i.e., allowing gas accretion to have arbitrary metallicities). Applying our satellite NE-CEM to EAGLE satellites at $\rscale{=}[0.2,0.4]$, we successfully recover their average infall time $\tauin{=}3.34{\pm}0.15\,\gyr$ and ICM-to-ISM oxygen fraction ratio $\ficm{=}0.23{\pm}0.02$. Our satellite NE-CEM model reveals that the constant contribution from enriched inflow is due to the rapid elevation of gas-phase metallicities by $(1-\figm)/(1-\ficm)$, where $\figm{=}0.16$ is the IGM-to-ISM oxygen fraction in EAGLE.

Taken together, our results provide a robust and rigorous physical interpretation of the complex SME profile observed in the DESI data and EAGLE simulation, serving as the foundation for our more comprehensive SME measurements and analyses in papers II and III. In the near future, with upcoming new data releases from DESI, our method will provide increasingly precise constraints on the role of environment in shaping galaxy chemical evolution across the cosmic web. Meanwhile, deeper spectroscopic surveys like the Prime Focus Spectrograph~\citep[PFS;][]{PFS2014} will enable SME measurements at higher redshifts than the DESI BGS, while spectroscopic cluster surveys like the Jiaotong University Spectroscopic Telescope~\citep[JUST;][]{JUST2024} will reveal an even clearer path of the dynamical and chemical evolution of satellite galaxies as they travel from the infall region to the violent core of clusters.

\section{Acknowledgments}

We thank Dirk Scholte, Rita Tojeiro, John Moustakas, Zheng Zheng, and
Zhongxu Zhai for their helpful comments and discussions. This work is
supported by the National Key Basic Research and Development Program of
China (No. 2023YFA1607800, 2023YFA1607804), the National Natural Science
Foundation of China (12595313, 12173024), and the China Manned Space
Program (No. CMS-CSST-2025-A04).  Y.L. is supported by the National Natural
Science Foundation of China~(123B2040) and the T.D. Lee scholarship. This
project is supported in part by Office of Science and Technology, Shanghai
Municipal Government (grant Nos. 24DX1400100, ZJ2023-ZD-001). Y.Z.
acknowledges the generous sponsorship from Yangyang Development Fund. Y.Z.
thanks Cathy Huang for her hospitality at the Zhangjiang High-tech Park.

This material is based upon work supported by the U.S. Department of Energy
(DOE), Office of Science, Office of High-Energy Physics, under Contract No.
DE–AC02–05CH11231, and by the National Energy Research Scientific Computing
Center, a DOE Office of Science User Facility under the same contract.
Additional support for DESI was provided by the U.S. National Science
Foundation (NSF), Division of Astronomical Sciences under Contract No.
AST-0950945 to the NSF’s National Optical-Infrared Astronomy Research
Laboratory; the Science and Technology Facilities Council of the United
Kingdom; the Gordon and Betty Moore Foundation; the Heising-Simons
Foundation; the French Alternative Energies and Atomic Energy Commission
(CEA); the National Council of Humanities, Science and Technology of Mexico
(CONAHCYT); the Ministry of Science, Innovation and Universities of Spain
(MICIU/AEI/10.13039/501100011033), and by the DESI Member Institutions:
\url{https://www.desi.lbl.gov/collaborating-institutions}. Any opinions,
findings, and conclusions or recommendations expressed in this material are
those of the author(s) and do not necessarily reflect the views of the U.
S. National Science Foundation, the U. S. Department of Energy, or any of
the listed funding agencies.

The authors are honored to be permitted to conduct scientific research on
I'oligam Du'ag (Kitt Peak), a mountain with particular significance to the
Tohono O’odham Nation.

\bibliographystyle{aasjournalv7}
\bibliography{sme_paper1.bib}



\end{document}